\documentclass[12pt,twoside, a4paper]{article}

\usepackage[dvips]{graphicx}
\usepackage{amssymb}
\usepackage{amssymb,amsmath}
\usepackage{graphicx}
\input{epsf.sty}  
\begin{document}
\begin{center}
\LARGE{\textbf{Vacua of $N=10$ three dimensional gauged
supergravity}}
\end{center}
\vspace{1 cm}
\textbf{Auttakit Chatrabhuti$^{1, \, 2}$} and \textbf{Parinya Karndumri$^{3,\, 4}$} \\
$^1$Theoretical High-Energy Physics and Cosmology Group, Department of Physics, \\
Faculty of Science, Chulalongkorn University, Bangkok 10330, Thailand\\
auttakit@sc.chula.ac.th \\
$^2$Thailand Center of Excellence in Physics, CHE, Ministry of Education, Bangkok 10400, Thailand\\
$^3$INFN, Sezione di Trieste, Italy\\
$^4$International School for Advanced Studies (SISSA), via Bonomea
265, 34136 Trieste, Italy\\
karndumr@sissa.it
\vspace{1 cm}
\begin{abstract}
We study scalar potentials and the corresponding vacua of $N=10$
three dimensional gauged supergravity. The theory contains 32 scalar
fields parametrizing the exceptional coset space
$\frac{E_{6(-14)}}{SO(10)\times U(1)}$. The admissible gauge groups
considered in this work involve both compact and non-compact gauge
groups which are maximal subgroups of $SO(10)\times U(1)$ and
$E_{6(-14)}$, respectively. These gauge groups are given by
$SO(p)\times SO(10-p)\times U(1)$ for $p=6,\ldots 10$, $SO(5)\times
SO(5)$, $SU(4,2)\times SU(2)$, $G_{2(-14)}\times SU(2,1)$ and
$F_{4(-20)}$. We find many AdS$_3$ critical points with various
unbroken gauge symmetries. The relevant background isometries
associated to the maximally supersymmetric critical points at which
all scalars vanish are also given. These correspond to the
superconformal symmetries of the dual conformal field theories in
two dimensions.
\end{abstract}
\newpage
\section{Introduction}
Gauged supergravities play an important role in many aspects of
string theory. Some of them arise as effective theories of string
compactifications in the presence of fluxes of various p-form
fields, see for example, \cite{gaugeSUGRA_flux} for a recent review.
Furthermore, they are very useful in the AdS/CFT correspondence
\cite{maldacena}. This is due to the fact that in gauged
supergravity theories, supersymmetry allows scalar potentials which
admit some critical points with negative cosmological constants, AdS
critical points. These critical points are of particular interest in
the context of the AdS/CFT correspondence because they correspond to
conformal field theories on the boundary of AdS space.
\\ \indent
In the original AdS$_5$/CFT$_4$ correspondence, critical points of
$N=8$ five dimensional gauged supergravity found in \cite{warner5D}
describe various phases of $N=4$ SYM. The correspondence is now
extended to other dimensions as well. These include AdS$_4$/CFT$_3$
and AdS$_3$/CFT$_2$ correspondences. The former is of interest in
the sense that it might give some insight to condensed matter
systems, for example, superconductors. Gauged supergravities in four
dimensions are useful to this study in much the same way as five
dimensional gauged supergravities in AdS$_5$/CFT$_4$. Vacua of $N=8$
four dimensional gauged supergravity have been classified in
\cite{warner, warner2} soon after its construction \cite{dewit 4D},
and recently, some new vacua of this theory have been identified in
\cite{fish2, Fisch_warner}. Although, a lot of works have been done
in finding critical points of this theory, it is expected that many
critical points remain to be found. On the other hand,
AdS$_3$/CFT$_2$ correspondence is a good place to test and study
many aspects of the AdS/CFT correspondence. This is because there
are many known two dimensional conformal field theories, and things
are more controllable in two dimensions. So, we hope to understand
AdS$_3$/CFT$_2$ in much more detail than the higher dimensional
analogues. In this case, three dimensional gauged supergravities
are, of course, the natural framework. In comparison with the higher
dimensional counterparts, AdS$_3$/CFT$_2$ is not only important for
understanding the AdS/CFT correspondence but also for the study of
black hole entropy, see \cite{krause lecture} for a review and
references therein.
\\ \indent Three dimensional Chern-Simons
gauged supergravity, see, for example, \cite{nicolai1, nicolai2,
nicolai3, N8} and \cite{dewit} for the construction, has a much
richer structure than the analogous theories in higher dimensions
due to the duality between vectors and scalars in three dimensions.
The admissible gauge groups include compact, non-compact,
non-semisimple and complex ones. Supersymmetry determines unique
scalar target spaces for theories with $N>8$, \cite{dewit1}. Some
works have been done in studying critical points or vacua of gauged
supergravities in three dimensions \cite{gkn, bs, AP, Fisch,
N16Vacua, fisch3}. The theories considered in these works have
$N=4,8,9,16$ supersymmetry, respectively. In this paper, we study
$N=10$ theory whose 32 scalar fields parametrize the coset
$\frac{E_{6(-14)}}{SO(10)\times U(1)}$. The admissible gauge groups
are subgroups of $E_{6(-14)}$. Some of the compact and non-compact
admissible gauge groups have been classified in \cite{dewit}. These
are gauge groups we will study in this work. The compact gauge
groups are $SO(p)\times SO(10-p)\times U(1)$ for $p=6,\ldots 10$ and
$SO(5)\times SO(5)$. The non-compact gauge groups are
$G_{2(-14)}\times SU(2,1)$, $SU(4,2)\times SU(2)$ and $F_{4(-20)}$.
All of these gauge groups are maximal subgroups of $SO(10)\times
U(1)$ and $E_{6(-14)}$, respectively.
\\ \indent
We will study some critical points of the scalar potentials in all
of the gaugings mentioned above by using the technique introduced in
\cite{warner}. In this ``subgroup method'', we start by choosing a
particular subgroup of the gauge group and study the potential on
the restricted scalar manifold which is invariant under this
subgroup. As a consequence of Schur's lemma, the critical points
found on this invariant manifold are critical points of the
potential on the whole scalar manifold, 32-dimensional
$\frac{E_{6(-14)}}{SO(10)\times U(1)}$ manifold in this work. This
method has been used to study critical points of scalar potentials
of $N=16$ gauged supergravity in \cite{N16Vacua} and in other
dimensions as well.
\\ \indent The paper is organized as follows. In section \ref{N10},
we review some useful ingredients to construct $N=10$ gauged
supergravity theory. We use the parametrization of the scalar coset
manifold $\frac{E_{6(-14)}}{SO(10)\times U(1)}$ in much the same way
as the $\frac{F_{4(-20)}}{SO(9)}$ coset in $N=9$ theory. All details
of the gauge group generators and other needed information can be
found in appendix \ref{detail}. Various vacua are given in section
\ref{N10vacua} including the background isometries of the maximally
supersymmetric critical points at which all scalars vanish. The
computations are carried out with the help of the computer program
\textsl{Mathematica} \cite{Mathematica}. We finally summarize our
results and give some conclusions in section \ref{conclusion}.
\section{$N=10$ three dimensional gauged supergravity}\label{N10}
In this section, we construct $N=10$ three dimensional gauged
supergravity using the formulation given in \cite{dewit}. The
procedure is essentially the same as that given in \cite{AP}, so we
will give only the needed ingredients and refer the reader to
\cite{dewit} for the full detail of the construction.
\\
\indent We start by giving a description of symmetric spaces. In
three dimensional gauged supergravity with $N>8$, scalar fields
parametrize a unique coset space of the form $G/H$. The group $G$
given by some non-compact real form of an exceptional group is the
global symmetry of the theory with the maximal compact subgroup $H$.
The subgroup $H$ is further decomposed to $SO(N)\times H'$ in which
$SO(N)$ is the R-symmetry. Note that the additional factor $H'$ does
not appear when $SO(N)$ is the maximal compact subgroup of $G$. This
is the case for $N=9$ and $N=16$ theories in which $G$ is given by
$F_{4(-20)}$ and $E_{8(8)}$, respectively. The $G$ generators
$t^{\mathcal{M}}$ decompose into $\{X^{IJ}, X^\alpha\}$ which are
generators of $\{SO(N), H'\}$ and non-compact generators $Y^A$.
\\ \indent
In general, the ungauged Lagrangian of the three dimensional
supergravity coupled to a non-linear sigma model is not invariant
under diffeomorphisms of the sigma model target space. In the
formulation of \cite{dewit}, the invariance of the Lagrangian is
constructed from some isometries of the target space including
appropriate field dependent $SO(N)$ transformations. The $G$
algebra, $\mathfrak{g}$, is then formed by the isometries of the
target space that can be extended to an invariance of the
Lagrangian. As shown in \cite{dewit}, under the map $\mathcal{V}$
\begin{equation}
\mathcal{V}: \mathfrak{g} \rightarrow \mathfrak{a}, \qquad
\mathcal{V}^{\mathcal{M}}_{\phantom{a}\mathcal{A}}t^{\mathcal{A}}=
\frac{1}{2}\mathcal{V}^{\mathcal{M}IJ}t^{IJ}+\mathcal{V}^\mathcal{M}_{\phantom{a}\alpha}t^\alpha+
\mathcal{V}^\mathcal{M}_{\phantom{a}A}t^A,
\end{equation}
the algebra $\mathfrak{g}$ is mapped to an associative subalgebra of
$\mathfrak{a}=\{t^{IJ}, t^\alpha, t^A\}$. The algebra $\mathfrak{a}$
is an extension of $SO(N)\times H'$ algebra, $\mathfrak{so}(N)\times
\mathfrak{h}'$, with the commutation relations given by
\begin{eqnarray}
[t^{IJ},t^{KL}]&=&-4\delta^{[I[K}t^{L]J]}, \qquad
[t^{IJ},t^A]=-\frac{1}{2}f^{IJ,AB}t_B,\qquad
[t^\alpha,t^\beta]=f^{\alpha
\beta}_{\phantom{as}\gamma}t^\gamma,\label{algebra}
\nonumber \\
\left
[t^{A},t^{B}\right]&=&\frac{1}{4}f^{AB}_{IJ}t^{IJ}+\frac{1}{8}C_{\alpha\beta}h^{\beta
AB}t^\alpha, \qquad [t^\alpha,t^A]=h^{\alpha
\phantom{a}A}_{\phantom{a}B}t^B
\end{eqnarray}
where $C_{\alpha\beta}$ and $h^{\alpha \phantom{a}A}_{\phantom{a}B}$
are an $H'$ invariant tensor and anti-symmetric tensors defined in
\cite{dewit}. $f^{IJ}_{ij}$ tensors are constructed from $N-1$
almost complex structures $f^P$, $p=2,\ldots N$. For symmetric
target spaces, all the $\mathcal{V}$'s are given by the expansion
\begin{equation}
L^{-1}t^\mathcal{M}L=\frac{1}{2}\mathcal{V}^{\mathcal{M}IJ}X^{IJ}+\mathcal{V}^\mathcal{M}_{\phantom{a}\alpha}X^\alpha+
\mathcal{V}^\mathcal{M}_{\phantom{a}A}Y^A,\label{cosetFormula}
\end{equation}
and the map $\mathcal{V}$ is now an isomorphism, see \cite{dewit}
for further detail. We have introduced ``flat'' indices $A,
B,\ldots$ for the scalar manifold. The target space metric $g_{ij}$,
$i,j=1,2,\ldots d=\textrm{dim}\, G/H$ is given by
\begin{equation}
g_{ij}=e^A_ie^B_j \delta_{AB}
\end{equation}
where the vielbein $e^A_i$ is encoded in the expansion
\begin{equation}
L^{-1} \partial_i L= \frac{1}{2}Q^{IJ}_i X^{IJ}+Q^\alpha_i
X^{\alpha}+e^A_i Y^A\, .\label{cosetFormula1}
\end{equation}
$Q^{IJ}_i$ and $Q^\alpha_i$ are composite connections for $SO(N)$
and $H'$, respectively. R-symmetry indices $I, J,\ldots =1,\ldots,
N$ and $\alpha, \beta,\ldots =1,\ldots, \textrm{dim} H'$. Finally,
the coset representative $L$ transforms under $G$ and $H$ by
multiplications from the left and right, respectively.
\\ \indent
The scalar manifold of $N=10$ theory is a 32 dimensional symmetric
space $\frac{E_{6(-14)}}{SO(10)\times U(1)}$. We will use the $E_6$
generators constructed in \cite{E6}. Notice that there is an
additional factor $H'=U(1)$ in this theory in contrast to $N=9$ and
$N=16$ theories studied in \cite{AP} and \cite{N16Vacua}. The 78
generators of $E_6$ are given in \cite{F4} for the first 52
generators and in \cite{E6} for the remaining 26. We can construct
the non-compact form $E_{6(-14)}$ by making 32 generators
non-compact using ``Weyl unitarity''. These transform as a spinor
representation of $SO(10)$ and are given by
\begin{equation}
Y^A=\Bigg\{
\begin{array}{rl}
ic_{A+21} &\textrm{for} \; A=1, \ldots, 8\\
ic_{A+28} &\textrm{for} \; A=9, \ldots, 16\\
ic_{A+37}&\textrm{for} \; A=17, \ldots, 32
\end{array} .
\end{equation}
The 46 compact generators are the generators of $SO(10)\times U(1)$
and are given in appendix \ref{detail}. The next ingredient we need
is the $f^{IJ}_{ij}$ tensors which can be read off from the second
commutator of \eqref{algebra} as we have described in \cite{AP}.
 \\ \indent We now come to various gaugings described by the
gauge invariant embedding tensor $\Theta_{\mathcal{M}\mathcal{N}}$.
This tensor acts as a projector on the symmetry group $G$ to the
gauge group $G_0$. The gauge generators are given by
\begin{equation}
J_{\mathcal{M}}=\Theta_{\mathcal{MN}}t^{\mathcal{N}}\, .
\end{equation}
The dimension of the gauge group is given by the rank of
$\Theta_{\mathcal{MN}}$. The requirement that these generators form
an algebra gives
\begin{equation}
\left[J_{\mathcal{M}},J_{\mathcal{N}}\right]=\hat{f}_{\mathcal{MN}}^{\phantom{assd}\mathcal{P}}J_{\mathcal{P}}\label{G0algebra}
\end{equation}
where $\hat{f}_{\mathcal{MN}}^{\phantom{assd}\mathcal{P}}$ are
structure constants of the gauge group. Using the $G$ algebra
$\left[t^{\mathcal{M}},t^{\mathcal{N}}\right]=f^{\mathcal{MN}}_{\phantom{asds}\mathcal{R}}t^{\mathcal{R}}$,
we can write \eqref{G0algebra} as
\begin{equation}
\Theta_{\mathcal{MP}}\Theta_{\mathcal{NQ}}f^{\mathcal{PQ}}_{\phantom{asds}\mathcal{R}}=\hat{f}_{\mathcal{MN}}^{\phantom{assd}\mathcal{P}}
\Theta_{\mathcal{PR}}\, .
\end{equation}
Together with the gauge invariant condition
$\hat{f}_{\mathcal{MN}}^{\phantom{assd}\mathcal{Q}}\Theta_{\mathcal{QP}}+\hat{f}_{\mathcal{MP}}^{\phantom{assd}\mathcal{Q}}
\Theta_{\mathcal{QN}}=0$, this implies the so-called quadratic
constraint
\begin{equation}
\Theta_{\mathcal{PL}}f^{\mathcal{KL}}_{\phantom{asds}\mathcal{(M}}\Theta_{\mathcal{N)K}}=0\,
.
\end{equation}
\indent From $\Theta_{\mathcal{M}\mathcal{N}}$, we can compute $A_1$
and $A_2$ tensors as well as the scalar potential via the so-called
T-tensors using
\begin{eqnarray}
A_1^{IJ}&=&-\frac{4}{N-2}T^{IM,JM}+\frac{2}{(N-1)(N-2)}\delta^{IJ}T^{MN,MN},\nonumber\\
A_{2j}^{IJ}&=&\frac{2}{N}T^{IJ}_{\phantom{as}j}+\frac{4}{N(N-2)}f^{M(I
m}_{\phantom{as}j}T^{J)M}_{\phantom{as}m}+\frac{2}{N(N-1)(N-2)}\delta^{IJ}f^{KL\phantom{a}m}_{\phantom{as}j}T^{KL}_{\phantom{as}m},
\nonumber \\
V&=&-\frac{4}{N}g^2(A_1^{IJ}A_1^{IJ}-\frac{1}{2}Ng^{ij}A_{2i}^{IJ}A_{2j}^{IJ})
\end{eqnarray}
with T-tensors
\begin{equation}
T_{\mathcal{A}\mathcal{B}}=\mathcal{V}^{\mathcal{M}}_{\phantom{a}\mathcal{A}}\Theta_{\mathcal{M}\mathcal{N}}
\mathcal{V}^{\mathcal{N}}_{\phantom{a}\mathcal{B}}\, .
\end{equation}
Supersymmetry imposes a projection constraint on $T^{IJ,KL}$
\begin{equation}
\mathbb{P}_{\boxplus}T^{IJ,KL}=0
\end{equation}
where $\boxplus$ denotes the representation $\boxplus$ of $SO(N)$.
For symmetric target spaces, it has been shown in \cite{dewit} that
the embedding tensor of the admissible gauge group must satisfy
\begin{equation}
\mathbb{P}_{R_0}\Theta_{\mathcal{MN}}=0\, .
\end{equation}
The representation $R_0$ of $G$ arises from decomposing the
symmetric product of two adjoint representations of $G$ under $G$.
Furthermore, the representation $R_0$, when branched under $SO(N)$,
is a unique representation in the above decomposition that contains
the $\boxplus$ representation of $SO(N)$.
\\ \indent
The embedding tensors for the compact gaugings with gauge groups
$SO(p)\times SO(10-p)\times U(1)$, $p=6,\ldots , 10$ and
$SO(5)\times SO(5)$ are given by \cite{dewit}
\begin{equation}
\Theta_{IJ,KL}=\theta
\delta^{KL}_{IJ}+\delta_{[I[K}\Xi_{L]J]}+\frac{1}{3}(5-p)\Theta_{U(1)}
\end{equation}
where
\begin{equation}
\Xi_{IJ}=\Bigg\{
\begin{array}{rl}
2\big(1-\frac{p}{10}\big)\delta_{IJ} & \textrm{for} \; I \leq p\\
-\frac{p}{5}\delta_{IJ} & \textrm{for} \; I > p
\end{array} , \qquad \theta=\frac{p-5}{5}\, .
\end{equation}
For $p=5$, the gauge group is $SO(5)\times SO(5)$ which lies
entirely in $SO(10)$. This is the case in which the $U(1)$ is not
gauged. The generators for these gauge groups can be obtained by
choosing appropriate generators of $SO(10)$, and the $U(1)$
generator is simply given by $2\tilde{c}_{70}$. We refer the reader
to appendix \ref{detail} for further details.
\\ \indent Non-compact gaugings considered in this work are those
given in \cite{dewit}. The gauge groups are $SU(4,2)\times SU(2)$,
$G_{2(-14)}\times SU(2,1)$ and $F_{4(-20)}$. We find the following
embedding tensors
\begin{eqnarray}
G_{2(-14)}\times SU(2,1)&:&
\Theta_{\mathcal{M}\mathcal{N}}=\eta^{G_2}_{\mathcal{M}\mathcal{N}}-\frac{2}{3}\eta^{SU(2,1)}_{\mathcal{M}\mathcal{N}}\label{ThetaG2}\\
SU(4,2)\times
SU(2)&:&\Theta_{\mathcal{M}\mathcal{N}}=\eta^{SU(4,2)}_{\mathcal{M}\mathcal{N}}-6\eta^{SU(2)}_{\mathcal{M}\mathcal{N}}\label{ThetaSp}\\
F_{4(-20)}&:&\Theta_{\mathcal{M}\mathcal{N}}=\eta^{F_{4(-20)}}_{\mathcal{M}\mathcal{N}}
\end{eqnarray}
where $\eta^{G_0}$ is the Cartan Killing form of the gauge group
$G_0$. The gauge generators of these three gaugings are given in
appendix \ref{detail}. \\ \indent We finally repeat the stationarity
condition for the critical points of the scalar potential
\cite{dewit}
\begin{equation}
3A_1^{IK}A^{KJ}_{2j}+Ng^{kl}A^{IK}_{2k}A^{KJ}_{3lj}=0\label{extremum}
\end{equation}
where $A^{KL}_{3lj}$ is defined by
\begin{equation}
A^{IJ}_{3ij}=\frac{1}{N^2}\bigg[-2D_{(i}D_{j)}A^{IJ}_1+g_{ij}A^{IJ}_1+A^{K[I}_1f^{J]K}_{ij}+2T_{ij}\delta^{IJ}-4D_{[i}T^{IJ}_{\phantom{as}
j]}-2T_{k[i}f^{IJk}_{\phantom{asd}j]}\bigg].
\end{equation}
For supersymmetric critical points, the unbroken supersymmetries are
encoded in the condition
\begin{equation}
A_1^{IK}A_1^{KJ}\epsilon^J=
-\frac{V_0}{4g^2}\epsilon^I=\frac{1}{N}(A_1^{IJ}A_1^{IJ}-\frac{1}{2}Ng^{ij}A_{2i}^{IJ}A_{2i}^{IJ})\epsilon^I\,
.\label{a1a2condition}
\end{equation}
The notations and all definitions are the same as those in
\cite{dewit}. In the next section, we will give the scalar potential
for each gauging along with the corresponding critical points.
\section{Vacua of $N=10$ gauged supergravity}\label{N10vacua}
In this section, we give some vacua of the $N=10$ gauged theory with
the gaugings described in the previous section. We will also discuss
the isometry groups of the background with maximal supersymmetry at
$L=\mathbf{I}$. This is a supersymmetric extension of the
$SO(2,2)\sim SO(1,2)\times SO(1,2)$ isometry group of AdS$_3$. A
similar study has been done in \cite{N16Vacua} and \cite{AP} for
$N=16$ and $N=9$ theories, respectively. For the full list of
superconformal groups in two dimensions, we refer the reader to
\cite{SC}. As a general strategy, we give the trivial critical point
in which all scalars are zero, $L=\mathbf{I}$, as the first critical
point. It is also useful to compare the cosmological constants of
other critical points with the trivial one. According to the AdS/CFT
correspondence, the cosmological constant $V_0$ is related to the
central charge in the dual CFT as $c\sim \frac{1}{\sqrt{-V_0}}$, so
we will give the ratio of the central charges for each non trivial
critical point with respect to the trivial critical point at
$L=\mathbf{I}$. We first start with compact gaugings.
\subsection{Vacua of compact gaugings}
The compact gauging includes gauge groups $SO(p)\times
SO(10-p)\times U(1)$ for $p=6,\ldots, 10$ and $SO(5)\times SO(5)$.
We give the scalar potential in $SO(p)\times SO(10-p)\times U(1)$
for $p=7,\ldots, 10$ gaugings in the $G_2$ invariant scalar sector.
For $SO(6)\times SO(4)\times U(1)$ gauging, we study the potential
in $SO(4)_{\textrm{diag}}$ and $SU(3)$ sectors. Finally, for
$SO(5)\times SO(5)$ gauging, we study the potential in
$SO(5)_{\textrm{diag}}$, $SO(4)_{\textrm{diag}}$ and
$SO(3)_{\textrm{diag}}$ sectors. All notations are the same as in
\cite{gkn} and \cite{AP}.
\subsubsection{$SO(10)\times U(1)$ gauging}
We will study the potential in the $G_2$ invariant scalar manifold.
From 32 scalars, there are four singlets under $G_2\subset SO(p)$,
$p=7,\ldots, 10$. These four scalars correspond to non-compact
directions of $SU(2,1)$. We use the same parametrization as in
\cite{N16Vacua}, namely using three compact generators of the
$SU(2)$ subgroup and one non-compact generator. With this
parametrization, the coset representative takes the form
\begin{equation}
L=e^{a_1c_{78}}e^{a_2\tilde{c}_{53}}e^{a_3c_{52}}e^{b_1(Y_1+Y_6)}e^{-a_3c_{52}}
e^{-a_2\tilde{c}_{53}}e^{-a_1c_{78}}\, .\label{G2coset}
\end{equation}
This choice of $L$ will also be used in the next three gauge groups.
In this $SO(10)\times U(1)$ gauging, the potential is given by
\begin{equation}
V=\frac{1}{2} g^2 [-101-28 \cosh(2 \sqrt{2} b_1)+\cosh(4 \sqrt{2}
b_1)].
\end{equation}
The potential does not depend on $a_1$, $a_2$ and $a_3$. \\ \indent
The first critical point is the trivial one in which all scalars are
zero. We find
\begin{equation}
V_0=-64g^2,\qquad A_1=-4\mathbf{I}_{10}\, .
\end{equation}
We use the notation $\mathbf{I}_n$ for an $n\times n$ identity
matrix from now on. This is the critical point with (10,0)
supersymmetry according to our convention. The corresponding
background isometry is $Osp(10|2,\mathbb{R})\times SO(2,1)$. \\
\indent The second critical point is at
$b_1=\frac{\cosh^{-1}2}{\sqrt{2}}$ with cosmological constant
$V_0=-100g^2$. This is a non-supersymmetric point. The ratio of the
central charges between this point and the maximally supersymmetric
point is
\begin{equation}
\frac{c_{(0)}}{c_{(1)}}=\sqrt{\frac{V_0^{(1)}}{V_0^{(0)}}}=\frac{5}{4}\,
.
\end{equation}
Here and from now on, the notations $c_{(0)}$ and $c_{(i)}$ mean the
central charges of the trivial and $i^{\textrm{th}}$ non trivial
critical points, respectively. \\ \indent For $a_1=a_3=0$, the coset
representative \eqref{G2coset} has a larger symmetry $SO(7)$. This
$SO(7)$ is embedded in $SO(8)$ in such a way that it stabilizes one
component of the $SO(8)$ spinor. In \cite{N16Vacua}, this $SO(7)$
has been called $SO(7)^{\pm}$ according to a component of
$\mathbf{8}_s$ or $\mathbf{8}_c$ is stabilized. Our critical point
is parametrized only by $b_1$, so has $SO(7)$ symmetry. Notice that
this point is very similar to the non-supersymmetric $SO(7)\times
SO(7)$ critical point of the $SO(8)\times SO(8)$ gauged $N=16$
theory given in \cite{N16Vacua} and the $SO(7)$ point in $SO(9)$
gauged $N=9$ theory studied in \cite{AP}. The similarity mentioned
here and in the followings means that the location and the value of
the cosmological constant relative to the trivial point are similar
for these points. We do not know whether this is only an accident or
there is a precise relation (to be specified if exists) between
these critical points.

\subsubsection{$SO(9)\times U(1)$ gauging}
The potential in this gauging is much more complicated than the
previous gauge group and depends on all four scalars. So, we use the
local $H=SO(10)\times U(1)$ symmetry to remove the $e^{-a_3c_{52}}
e^{-a_2\tilde{c}_{53}}e^{-a_1c_{78}}$ factor in \eqref{G2coset} to
simplify the computation and reduce the calculation time. The
potential is given in appendix \ref{SO9_potential}. Although we do
not have a systematic way of finding critical points of this
complicated potential, we find some critical points, numerically.
\\ \indent The first critical point is the maximally supersymmetric
(9,1) point
\begin{eqnarray}
a_1&=&a_2=a_3=b_1=0,\qquad \qquad V_0=-64g^2, \nonumber \\
A_1&=&\textrm{diag}\left(-4,-4,-4,-4,-4,-4,-4,-4,-4,4\right).
\end{eqnarray}
The background isometry is given by $Osp(9|2, \mathbb{R})\times
Osp(1|2, \mathbb{R})$. \\ \indent The second critical point is given
by
\begin{eqnarray}
b_1&=&\frac{1}{\sqrt{2}}\cosh^{-1}\frac{7}{3},\qquad a_1=\pi,\,\,
a_2=\frac{3\pi}{2},\,\, a_3=\frac{\pi}{2},\,\,\,
V_0=-\frac{1024}{9}g^2, \nonumber \\
A_1&=&\textrm{diag}\left(-8,-8,-8,-8,-8,-8,-8,\frac{16}{3},-\frac{16}{3},-\frac{16}{3}\right).
\end{eqnarray}
This $G_2$ critical point has (2,1) supersymmetry with
\begin{equation}
\frac{c_{(0)}}{c_{(1)}}=\frac{4}{3}.
\end{equation}
This critical point should be compared with the (1,1) $G_2\times
G_2$ point in the $SO(8)\times SO(8)$ gauged $N=16$ theory. The two
points have similar locations and values of the cosmological
constant relative to the trivial point.
\\ \indent The third
critical point in this gauging is given by
\begin{eqnarray}
b_1&=&\frac{1}{\sqrt{2}}\cosh^{-1}2,\qquad
a_1=a_3=\frac{\pi}{2}, \qquad a_2=\textrm{arbitrary},\qquad V_0=-100g^2,\nonumber \\
A_1&=&\textrm{diag}\left(-7,-7,-7,-7,-7,-7,-7,-7,7,-5\right).
\end{eqnarray}
This is a (1,0) point with $G_2$ symmetry and
\begin{equation}
\frac{c_{(0)}}{c_{(2)}}=\frac{5}{4}.
\end{equation}

\subsubsection{$SO(8)\times SO(2)\times U(1)$ gauging}
The potential in the $G_2$ sector is given by
\begin{eqnarray}
V&=&\frac{1}{4096}e^{-4 \sqrt{2} b_1} g^2 [3 (-1+e^{\sqrt{2} b_1})^8
\cos(4 a_1)+4 (-1+e^{\sqrt{2} b_1})^6 \cos(2 a_1) [27\nonumber \\ &
&+170 e^{\sqrt{2} b_1}+27 e^{2 \sqrt{2} b_1}+4 (e^{\sqrt{2}
b_1}-1)^2 \cos^2a_1  \cos(2 a_3)]\nonumber \\ & &+8 (e^{\sqrt{2}
b_1}-1)^6 \cos^2a_1 [2 (13+86 e^{\sqrt{2} b_1 }+13 e^{2 \sqrt{2}
b_1}) \cos(2 a_3)\nonumber \\ & &+(e^{\sqrt{2} b_1}-1)^2 \cos^2a_1
\cos(4 a_3)]-2 e^{4 \sqrt{2} b_1} [88549+21112 \cosh(\sqrt{2}
b_1)\nonumber \\ & &+22148 \cosh(2 \sqrt{2} b_1)-56 \cosh(3 \sqrt{2}
b_1)-681 \cosh(4 \sqrt{2} b_1)]].
\end{eqnarray}
The potential does not depend on $a_2$.
We find the following critical points. \\
\indent First of all, when $a_1=a_2=a_3=b_1=0$, we find the
maximally supersymmetric critical points. At this point, we find
\begin{eqnarray}
V_0&=&-64g^2,\nonumber \\
A_1&=&\textrm{diag}\left(-4,-4,-4,-4,-4,-4,-4,-4,4,4\right).
\end{eqnarray}
This point has (8,2) supersymmetry and $Osp(8|2,\mathbb{R})\times
Osp(2|2, \mathbb{R})$ as the background isometry group. \\ \indent
The next point is given by
\begin{equation}
b_1=\cosh^{-1}2,\qquad a_1=a_3=0,\qquad V_0=-100g^2\, .
\end{equation}
This is an $SO(7)$ non-supersymmetric point with
\begin{equation}
\frac{c_{(0)}}{c_{(1)}}=\frac{5}{4}.
\end{equation}
This point is very similar to the non-supersymmetric $SO(7)\times
SO(7)$ point of the $SO(8)\times SO(8)$ gauged $N=16$ theory studied
in \cite{N16Vacua}.
 \\
\indent The third critical point is given by
\begin{eqnarray}
b_1&=&\frac{1}{\sqrt{2}}\cosh^{-1} \frac{7}{3},\qquad a_1=0,\qquad
a_3=\frac{\pi}{2},\qquad V_0=-\frac{1024}{9}g^2,\nonumber
\\
A_1&=&\left(
\begin{array}{cccccccccc}
 -8 & 0 & 0 & 0 & 0 & 0 & 0 & 0 & 0 & 0 \\
 0 & -8 & 0 & 0 & 0 & 0 & 0 & 0 & 0 & 0 \\
 0 & 0 & -8 & 0 & 0 & 0 & 0 & 0 & 0 & 0 \\
 0 & 0 & 0 & -8 & 0 & 0 & 0 & 0 & 0 & 0 \\
 0 & 0 & 0 & 0 & -8 & 0 & 0 & 0 & 0 & 0 \\
 0 & 0 & 0 & 0 & 0 & -8 & 0 & 0 & 0 & 0 \\
 0 & 0 & 0 & 0 & 0 & 0 & -8 & 0 & 0 & 0 \\
 0 & 0 & 0 & 0 & 0 & 0 & 0 & -\frac{16}{3} & 0 & 0 \\
 0 & 0 & 0 & 0 & 0 & 0 & 0 & 0 & x_1 & x_2 \\
 0 & 0 & 0 & 0 & 0 & 0 & 0 & 0 & x_2 & x_3
\end{array}
\right)
\end{eqnarray}
where
\begin{eqnarray}
x_1&=&-\frac{4}{3} [-5+\cos(2 a_2)], \qquad x_2=\frac{4}{3} \sin(2a_2), \nonumber \\
x_3&=&\frac{4}{3} [5+\cos(2a_2)].
\end{eqnarray}
We find that this is the (1,1) point with $G_2$ symmetry, and the
diagonalized $A_1$ tensor is given by
\begin{equation}
A_1=\textrm{diag}\left(-8,-8,-8,-8,-8,-8,-8,8,-\frac{16}{3},\frac{16}{3}\right).
\end{equation}
The ratio of the central charges is
\begin{equation}
\frac{c_{(0)}}{c_{(2)}}=\frac{4}{3}\, .
\end{equation}
This point is similar to the $G_2\times G_2$ point with (1,1)
supersymmetry in $SO(8)\times SO(8)$ gauged $N=16$ theory.

\subsubsection{$SO(7)\times SO(3)\times U(1)$ gauging}
In this gauging, we still work with the $G_2$ invariant scalar
sector. The potential is given by
\begin{equation}
V=-\frac{1}{32} g^2 [1301+448 \cosh(\sqrt{2} b_1)+308 \cosh(2
\sqrt{2} b_1)-9 \cosh(4 \sqrt{2} b_1)].
\end{equation}
This case is very similar to the $SO(10)\times U(1)$ gauging in the
sense that the potential dose not depend on $a_1$, $a_2$ and $a_3$
and admits two critical points. \\ \indent The first critical point
is as usual at $L=\textbf{I}$. This point is a (7,3) point with
\begin{eqnarray}
V_0&=&-64g^2 \nonumber \\
A_1&=&\textrm{diag}(-4,-4,-4,-4,-4,-4,-4,4,4,4).
\end{eqnarray}
The background isometry is $Osp(7|2,\mathbb{R})\times
Osp(3|2,\mathbb{R})$.
\\ \indent The second critical point is given
by
\begin{equation}
b_1=\frac{1}{\sqrt{2}}\cosh^{-1}\frac{7}{3}, \qquad
V_0=-\frac{1024}{9}g^2\, .
\end{equation}
The $A_1$ tensor is very complicated, so we give its explicit form
in appendix \ref{A1tensor} equation \eqref{A1SO7}. Remarkably, the
complicated matrix $M_3^{(1)}$ can be diagonalized to
$\textrm{diag}\left(8,8,\frac{16}{3}\right)$. This gives
\begin{equation}
A_1=\textrm{diag}\left(-8,-8,-8,-8,-8,-8,-8,8,8,\frac{16}{3}\right).
\end{equation}
So, this critical point has (0,1) supersymmetry with
\begin{equation}
\frac{c_{(0)}}{c_{(1)}}=\frac{4}{3}\, .
\end{equation}
Notice that this point has $G_2$ symmetry although it is
characterized only by $b_1$. This is because the $SO(7)$ in the
gauge group is not the same as $SO(7)^{\pm}$, and $b_1$ is not
invariant under this $SO(7)$. The $SO(7)$ in the gauge group is
embedded in $SO(8)$ as $\mathbf{8}_v\rightarrow
\mathbf{7}+\mathbf{1}$. This point is similar to the (1,1)
$G_2\times G_2$ point in \cite{N16Vacua}.

\subsubsection{$SO(6)\times SO(4)\times U(1)$ gauging}
We first study the potential in the $SO(4)_{\textrm{diag}}$ scalar
sector. There are four singlets in this sector corresponding the
non-compact directions of $SO(2,2)\sim SO(2,1)\times SO(2,1)$. We
parametrize the coset representative by
\begin{equation}
L=e^{a_1[V_1,V_2]}e^{b_1V_1}e^{-a_1[V_1,V_2]}e^{a_2[V_3,V_4]}e^{b_2V_1}e^{-a_2[V_3,V_4]},\label{LSO4}
\end{equation}
where
\begin{eqnarray}
V_1&=&j_1+j_2,\nonumber \\
V_2&=&j_3-j_4,\nonumber \\
V_3&=&j_3+j_4,\nonumber \\
V_4&=&j_1-j_2,
\end{eqnarray}
and
\begin{eqnarray}
j_1&=&Y_1+Y_5-Y_9+Y_{13}-Y_{17}-Y_{21}+Y_{30}+Y_{32},\nonumber \\
j_2&=&Y_2+Y_{10}-Y_{11}+Y_{18}+Y_{19}-Y_{28}+Y_{31}+Y_3,\nonumber \\
j_3&=&Y_4+Y_7+Y_{12}-Y_{15}+Y_{20}+Y_{23}+Y_{26}-Y_{27},\nonumber \\
j_4&=&Y_6-Y_8+Y_{14}+Y_{16}-Y_{22}+Y_{24}+Y_{25}-Y_{29}\, .
\end{eqnarray}
We find the potential
\begin{eqnarray}
V&=&-2 e^{-4 \sqrt{2} (b_1+b_2)} [1+4 e^{4 \sqrt{2} b_1}+e^{8
\sqrt{2} b_1}+4 e^{4 \sqrt{2} b_2}+e^{8 \sqrt{2} b_2}\nonumber \\ &
&+12 e^{4 \sqrt{2} (b_1+b_2)}+e^{8 \sqrt{2} (b_1+b_2)}+4 e^{4
\sqrt{2} (2 b_1+b_2)}+4 e^{4 \sqrt{2} (b_1+2 b_2)}] g^2\, .
\end{eqnarray}
There is no non-trivial critical point in this potential. So, there
is no critical point with $SO(4)_{\textrm{diag}}$ symmetry. \\
\indent Next, we will consider the $SU(3)$ invariant sector. The
$SU(3)$ is a subgroup of $SO(6)\sim SU(4)$. There are eight singlets
in this sector. The coset representative is parametrized by
\begin{equation}
L=e^{a_1c_{36}}e^{a_2c_{51}}e^{a_3c_{52}}e^{a_4\tilde{c}_{53}}e^{a_5c_{77}}e^{a_6c_{78}}e^{b_1Y_1}e^{b_2Y_3}
\end{equation}
in which the eight scalars correspond to non-compact directions of
$SU(2,2)$. As usual, we have used the local $H$ symmetry to simplify
the parametrization of $L$. The potential is given in appendix
\ref{SO6potential}. We find two critical points. \\ \indent The
trivial (6,4) critical point at $L=\mathbf{I}$ is given by
\begin{eqnarray}
V_0&=&-64g^2,\nonumber \\
A_1&=&\textrm{diag}\left(-4,-4,-4,-4,-4,-4,4,4,4,4 \right).
\end{eqnarray}
The background isometry is $Osp(6|2, \mathbb{R})\times Osp(4|2,
\mathbb{R})$.
\\ \indent The non trivial critical point is given by
\begin{eqnarray}
a_i&=&\frac{\pi}{2}, \qquad i=1,\ldots , 6 ,\nonumber \\
b_1&=&b_2=\cosh^{-1}\sqrt{3},\qquad V_0=-144g^2, \nonumber \\
A_1&=&\textrm{diag}\left(-10,-10,-10,-10,-10,-10,6,6,10,10 \right).
\end{eqnarray}
This point preserves (0,2) supersymmetry and $SU(3)$ symmetry. The
ratio of the central charges is
\begin{equation}
\frac{c_{(0)}}{c_{(1)}}=\frac{3}{2}\, .
\end{equation}

\subsubsection{$SO(5)\times SO(5)$ gauging}
We start with the potential in the $SO(5)_{\textrm{diag}}$ scalar
sector. There are two singlets in this sector corresponding to the
non-compact directions of $SL(2)$. We parametrize the coset
representative by
\begin{equation}
L=e^{a_1V}e^{b_1U}e^{-a_1V}
\end{equation}
where the compact and non-compact generators of $SL(2)$ are given by
\begin{eqnarray}
V&=&\frac{1}{\sqrt{2}}\left(c_{11}-c_{17}+c_{32}-c_{48}+c_{75}+\frac{\sqrt{3}}{2}\tilde{c}_{70}\right),\\
U&=&Y_3-Y_5-Y_{12}+Y_{16}+Y_{17}-Y_{18}+Y_{27}+Y_{29}\, .
\end{eqnarray}
The potential is given by
\begin{equation}
V=-8g^2(5+3\cosh (4b_1))
\end{equation}
which does not have any non-trivial critical points. \\
\indent We then move to smaller unbroken gauge symmetry namely
$SO(4)_{\textrm{diag}}$. The parametrization of $L$ is the same as
in \eqref{LSO4}. The potential turns out to be the same as that of
$SO(6)\times SO(4)\times U(1)$ gauging, and, of course, does not
have any non trivial critical points.
\\ \indent To proceed further, we
need to reduce the residual symmetry to a smaller group. The next
sector we will consider is $SO(3)_{\textrm{diag}}$. There are eight
singlets in this sector. These are non-compact directions of
$SO(4,2)\sim SU(2,2)$. We parametrize the coset representative in
this sector by
\begin{equation}
L=e^{a_1c_{10}}e^{a_2c_{14}}e^{a_3c_{15}}e^{a_4c_{19}}e^{a_5c_{20}}e^{a_6c_{21}}e^{b_1Z_1}e^{b_2Z_2}
\end{equation}
where
\begin{equation}
Z_1=Y_1+Y_{11}-Y_{20}-Y_{29},\qquad Z_2=Y_2+Y_{13}-Y_{24}+Y_{27}\, .
\end{equation}
The potential depends on all eight scalars. Its explicit form is
given in appendix \ref{SO5potential}. \\ \indent The trivial (5,5)
critical point at $L=\mathbf{I}$ is characterized by
\begin{equation}
V_0=-64g^2,\qquad
A_1=\textrm{diag}\left(-4,-4,-4,-4,-4,4,4,4,4,4\right).
\end{equation}
The corresponding background isometry group is
$Osp(5|2,\mathbb{R})\times Osp(5|2,\mathbb{R})$.
\\ \indent
We find a non trivial critical point given by
\begin{eqnarray}
a_i&=&\frac{\pi}{2}, \,\,\, i=1,\ldots, 6,\qquad b_2=0, \nonumber \\
b_1&=&\frac{\cosh^{-1}5}{2}, \qquad V_0=-256g^2,\nonumber \\
A_1&=&\textrm{diag}\left(-8,-8,-8,16,16,-16,-16,16,16,16\right).
\end{eqnarray}
This critical point has (3,0) supersymmetry with the ratio of the
central charges
\begin{equation}
\frac{c_{(0)}}{c_{(1)}}=2\, .
\end{equation}

\subsection{Vacua of non-compact gaugings}
We now consider non-compact gaugings with gauge groups
$SU(4,2)\times SU(2)$, $G_{2(-14)}\times SU(2,1)$ and $F_{4(-20)}$.
At $L=\mathbf{I}$, the gauge group is broken down to its maximal
compact subgroup, and the bosonic part of the background isometry is
formed by this subgroup and $SO(2,2)$. These three gauge groups
contain $SU(3)$ subgroup, so we study the potential in the $SU(3)$
scalar sector in all non-compact gaugings. For $G_{2(-14)}\times
SU(2,1)$ and $F_{4(-20)}$ gaugings, the $SU(3)\subset G_2$ sector
consists of eight scalars which is twice the number of scalars in
the $G_2$ sector. The $SU(3)$ is embedded in $G_2$ as
$\mathbf{7}\rightarrow \mathbf{3}+\bar{\mathbf{3}}+\mathbf{1}$. The
eight scalars correspond to non-compact directions of the
$SO(4,2)\sim SU(2,2)\subset E_{6(-14)}$. For $SU(4,2)\times SU(2)$
gauging, the $SU(3)$ is embedded in $SU(4)\subset SU(4,2)$ as
$\mathbf{4}\rightarrow \mathbf{3}+\mathbf{1}$. Similarly, the eight
scalars are described by non-compact directions of $SU(2,2)$. This
sector is essentially the same as that used in $SO(6)\times
SO(4)\times U(1)$ gauging.
\\ \indent Fortunately, we do not
need to deal with all eight scalars. In these three gaugings, four
of the eight $SU(3)$ singlets lie along the gauge group, so only
four directions orthogonal to the gauge group are relevant. This is
because the singlets which are parts of the gauge group will drop
out from the potential and correspond to flat directions of the
potential. The relevant four singlets are contained in the $SU(2,1)$
sub group of $SU(2,2)$. We also study the potentials in other
sectors specific to each gauging. The details of these sectors will
be explained below.

\subsubsection{$G_{2(-14)}\times SU(2,1)$ gauging}
If we study the potential in the $G_2$ sector in this gauging, we
will find the constant potential. This is because all scalars in the
$G_2$ sector are parts of the gauge group and will drop out from the
potential. We then start with $SU(3)\subset G_2$ sector. As
discussed above, this sector contains four relevant scalars
parametrized by
\begin{equation}
L=e^{a_1c_{52}}e^{a_2c_{78}}e^{a_3\tilde{c}_{53}}e^{b_1(Y_1-Y_6)}e^{-a_3\tilde{c}_{53}}e^{-a_2c_{78}}e^{-a_1c_{52}}\,
.
\end{equation}
The potential is given by
\begin{equation}
V=\frac{1}{18} g^2 [-101-28 \cosh(2 \sqrt{2} b_1)+\cosh(4 \sqrt{2}
b_1)].
\end{equation}
There are two critical points. The first one is the trivial critical
point given by $L=\mathbf{I}$ and
\begin{eqnarray}
V_0&=&-\frac{64}{9}g^2,\nonumber\\
A_1&=&\textrm{diag}\left(-\frac{4}{3},-\frac{4}{3},-\frac{4}{3},-\frac{4}{3},-\frac{4}{3},-\frac{4}{3},-\frac{4}{3}
,\frac{4}{3},\frac{4}{3},\frac{4}{3}\right).
\end{eqnarray}
We find that this point has (7,3) supersymmetry. The symmetry of
this point is given by the maximal compact subgroup $G_2\times
SU(2)\times U(1)$ of $G_{2(-14)}\times SU(2,1)$. The left handed
supercharges transform as $\mathbf{7}$ under $G_2$ while the right
handed supercharges transform as $\mathbf{3}$ under the $SU(2)\sim
SO(3)$. So, the background isometry is given by $G(3)\times
Osp(3|2,\mathbb{R})$. \\ \indent The second critical point is
characterized by
\begin{eqnarray}
b_1&=&\frac{\cosh^{-1}2}{\sqrt{2}},\qquad
V_0=-\frac{100}{9}g^2,\nonumber \\
A_1&=&\left(
\begin{array}{cccccccccc}
 -\frac{7}{3} & 0 & 0 & 0 & 0 & 0 & 0 & 0 & 0 & 0 \\
 0 & -\frac{7}{3} & 0 & 0 & 0 & 0 & 0 & 0 & 0 & 0 \\
 0 & 0 & -\frac{7}{3} & 0 & 0 & 0 & 0 & 0 & 0 & 0 \\
 0 & 0 & 0 & -\frac{7}{3} & 0 & 0 & 0 & 0 & 0 & 0 \\
 0 & 0 & 0 & 0 & -\frac{11}{3} & 0 & 0 & 0 & 0 & 0 \\
 0 & 0 & 0 & 0 & 0 & -\frac{7}{3} & 0 & 0 & 0 & 0 \\
 0 & 0 & 0 & 0 & 0 & 0 & -\frac{7}{3} & 0 & 0 & 0 \\
 0 & 0 & 0 & 0 & 0 & 0 & 0 & y_1& y_4 & y_5 \\
 0 & 0 & 0 & 0 & 0 & 0 & 0 & y_4 & y_2 & y_6 \\
 0 & 0 & 0 & 0 & 0 & 0 & 0 & y_5 & y_6 & y_3
\end{array}
\right)
\end{eqnarray}
where
\begin{eqnarray}
y_1&=&\frac{1}{6} [13-\cos(2 a_1)-2 \cos^2a_1 \cos(2 a_2)],\nonumber
\\
y_2&=&\frac{1}{6} [13+\cos(2 a_1)-2 \cos(2 a_2) \sin^2a_1 ], \nonumber \\
y_3&=&\frac{1}{3} (6+\cos(2 a_2)) ,\qquad y_4=\frac{1}{3} \cos^2a_2 \sin(2 a_1), \nonumber \\
y_5&=&-\frac{1}{3} \cos a_1  \sin(2 a_2) ,\qquad y_6=\frac{1}{3}
\sin a_1  \sin(2 a_2) \, .
\end{eqnarray}
We can diagonalize $A_1$ to
\begin{equation}
A_1=\textrm{diag}\left(-\frac{11}{3},-\frac{7}{3},-\frac{7}{3},-\frac{7}{3},-\frac{7}{3},
-\frac{7}{3},-\frac{7}{3},\frac{7}{3},\frac{7}{3},\frac{5}{3}\right)
\end{equation}
from which we find that this is a (0,1) supersymmetric critical
point. The ratio of the central charges relative to the
$L=\mathbf{I}$ point is
\begin{equation}
\frac{c_{(0)}}{c_{(1)}}=\frac{5}{4}\, .
\end{equation}
This $SU(3)$ point is closely related to the (0,1) $SU(3)$ point in
$G_{2(-14)}\times SL(2)$ gauged $N=9$ theory in \cite{AP}.
\\ \indent We now study the potential in different sector,
$SU(2)_{\textrm{diag}}$ sector. From the $SU(3)$ sector discussed
above, the next symmetry to consider could be the $SU(2)\subset
SU(3)$. In general, we expect more scalars than those appearing in
the $SU(3)$ sector. This will make the calculation takes much longer
time. We then consider $SU(2)_{\textrm{diag}}$ sector in which
$SU(2)_{\textrm{diag}}\subset SU(2)\times SU(2)$. The first and
second $SU(2)$'s are subgroups of $SU(3)\subset G_{2(-14)}$ and
$SU(2,1)$, respectively. There are four singlets in this sector
corresponding to the non-compact directions of $SO(4,1) \sim
Sp(1,1)$. We choose to parametrize the coset representative by
applying three $SO(3)\subset SO(4)\sim SO(3)\times SO(3)$ rotations
as follow
\begin{equation}
L=e^{a_1c_{8}}e^{a_2c_{17}}e^{a_3c_{20}}e^{b_1(Y_2-Y_{16}+Y_{19}+Y_{29})}e^{-a_3c_{20}}e^{-a_2c_{17}}e^{-a_1c_{8}}\,
.
\end{equation}
The potential is
\begin{equation}
V=\frac{1}{72} g^2 [-269-192 \cosh(2 b_1)-52 \cosh(4 b_1)+\cosh(8
b_1)]\, .
\end{equation}
There is one non trivial critical points given by
\begin{equation}
b_1=\cosh^{-1}\sqrt{2}, \qquad V_0=-16g^2\, .
\end{equation}
This is a supersymmetric point with the associated $A_1$ tensor
given in appendix \ref{A1tensor} equation \eqref{G2SU2d}. After
diagonalization, we find
\begin{equation}
A_1=\textrm{diag}\left(-4,-4,-4,-4,-\frac{10}{3},-2,-2,2,2,2\right)
\end{equation}
which gives (2,3) supersymmetry. The ratio of the central charges is
\begin{equation}
\frac{c_{(0)}}{c_{(2)}}=\frac{3}{2}\, .
\end{equation}
This critical point has $SU(2)_{\textrm{diag}}\times U(1)$ symmetry.

\subsubsection{$F_{4(-20)}$ gauging}
In this gauging with simple gauge group, we study the potential in
the $G_2$ and $SU(3)$ scalar sectors. We start with the $G_2$
sector. Two of the four scalars are parts of the gauge group, so we
only need to parametrize the coset representative with the other two
scalars. These two scalars correspond to the non-compact directions
of $SL(2)$. The $L$ is then parametrized by
\begin{equation}
L=e^{a_1c_{52}}e^{b_1(Y_{25}+Y_{30})}e^{-a_1c_{52}}\, .
\end{equation}
The potential is
\begin{equation}
V=\frac{g^2}{8}[-101-28\cosh(2\sqrt{2}b_1)+\cosh(4\sqrt{2}b_1)].
\end{equation}
There are two critical points. The first one is trivial and given by
\begin{eqnarray}
L&=&\mathbf{I}, \qquad V_0=-16g^2, \nonumber \\
A_1&=&\textrm{diag}\left(-2,-2,-2,-2,-2,-2,-2,-2,-2,2\right).
\end{eqnarray}
This is the maximally supersymmetric point with (9,1) supersymmetry.
The gauge symmetry is broken down to its maximal compact subgroup
$SO(9)$, and the background isometry is $Osp(9|2, \mathbb{R})\times
Osp(1|2, \mathbb{R})$. \\ \indent The second critical point is given
by
\begin{eqnarray}
b_1&=&\frac{\cosh^{-1}2}{\sqrt 2},\qquad V_0=-25g^2, \nonumber \\
A_1&=&\left(
\begin{array}{cccccccccc}
 -\frac{7}{2} & 0 & 0 & 0 & 0 & 0 & 0 & 0 & 0 & 0 \\
 0 & -\frac{7}{2} & 0 & 0 & 0 & 0 & 0 & 0 & 0 & 0 \\
 0 & 0 & -\frac{7}{2} & 0 & 0 & 0 & 0 & 0 & 0 & 0 \\
 0 & 0 & 0 & -\frac{7}{2} & 0 & 0 & 0 & 0 & 0 & 0 \\
 0 & 0 & 0 & 0 & -\frac{7}{2} & 0 & 0 & 0 & 0 & 0 \\
 0 & 0 & 0 & 0 & 0 & -\frac{7}{2} & 0 & 0 & 0 & 0 \\
 0 & 0 & 0 & 0 & 0 & 0 & -\frac{7}{2} & 0 & 0 & 0 \\
 0 & 0 & 0 & 0 & 0 & 0 & 0 & w_1 & w_3  & 0 \\
 0 & 0 & 0 & 0 & 0 & 0 & 0 & w_3  & w_2 & 0 \\
 0 & 0 & 0 & 0 & 0 & 0 & 0 & 0 & 0 & \frac{11}{2}
\end{array}
\right)
\end{eqnarray}
where
\begin{equation}
w_1=-3-\frac{1}{2} \cos(2 a_1),\,\,\,w_2=\frac{1}{2} [-6+\cos(2
a_1)],\,\,\,w_3=\cos a_1  \sin a_1\,.
\end{equation}
The $A_1$ tensor can be diagonalized to
\begin{equation}
A_1=\textrm{diag}\left(\frac{11}{2},-\frac{7}{2},-\frac{7}{2},-\frac{7}{2},
-\frac{7}{2},-\frac{7}{2},-\frac{7}{2},-\frac{7}{2},-\frac{7}{2},-\frac{5}{2}\right).\label{A1G2}
\end{equation}
This critical point is a (1,0) point with
\begin{equation}
\frac{c_{(0)}}{c_{(1)}}=\frac{5}{4}
\end{equation}
and preserves $SO(7)\subset SO(9)\subset F_{4(-20)}$ symmetry.\\
\indent In the $SU(3)$ sector, there are eight singlets, but four of
them are parts of the $F_{4(-20)}$. So, there are four singlets
orthogonal to the gauge group. These are non-compact directions of
$SU(2,1)$, and $L$ can be parametrized by
\begin{equation}
L=e^{a_1c_{34}}e^{a_2c_{49}}e^{a_3c_{52}}e^{b_1Y_{21}}e^{-a_3c_{52}}e^{-a_2c_{49}}e^{-a_1c_{34}}\,
.
\end{equation}
The potential is given by
\begin{equation}
V=\frac{g^2}{8}[-101-28\cosh(2\sqrt{2}b_1)+\cosh(4\sqrt{2}b_1)]
\end{equation}
which is the same as the potential in the $G_2$ sector. The
non-trivial critical point is at the same position and cosmological
constant, $b_1=\cosh^{-1} 2$, $V_0=-25g^2$. The residual symmetry is
$SO(7)$ as in the previous critical point. Although the $A_1$ tensor
in this case is more complicated, it is the same as \eqref{A1G2}
after diagonalization. The explicit form of $A_1$ is given in
appendix \ref{A1tensor} equation \eqref{A1F4SU3}.

\subsubsection{$SU(4,2)\times SU(2)$ gauging}
This gauging is the most difficult one to find a suitable scalar
sector in order to reveal non trivial critical points and still have
a manageable number of scalars. We start with the
$SO(4)_\textrm{diag}$ scalar sector. The $SO(4)_\textrm{diag}$ is
formed by taking the subgroup $SU(2)\times SU(2)\times SU(2)\times
SU(2)$ of $SU(4,2)\times SU(2)$. The first two $SU(2)$'s are
subgroups of $SU(4)\subset SU(4,2)$, the third $SU(2)$ is the
$SU(2)\subset SU(4,2)$. Our $SO(4)_\textrm{diag}$ is the diagonal
subgroup of $(SU(2)\times SU(2))\times (SU(2)\times SU(2))\sim
SO(4)\times SO(4)$. There are two singlets in this sector. These are
non-compact directions of $SL(2)$, and $L$ can be parametrized by
\begin{eqnarray}
L&=&e^{a_1c_{15}}e^{b_1\tilde{Y}}e^{-a_1c_{15}}, \nonumber \\
\tilde{Y}&=&Y_1+Y_2-Y_6-Y_7-Y_9+Y_{10}-Y_{14}+Y_{15} \nonumber \\
& &+Y_{17}-Y_{18}-Y_{22}+Y_{23}-Y_{27}+Y_{28}-Y_{29}-Y_{32}
\end{eqnarray}
which, unfortunately, gives a constant potential $V=-16g^2$. So, we
move to a smaller residual symmetry to obtain a non trivial
structure of the potential.
\\ \indent
We now study the potential in the scalar sector parametrizing the
$SU(3)$ invariant manifold. This $SU(3)$ is a subgroup of
$SU(4)\subset SU(4,2)$. The eight singlet scalars in this sector are
the non-compact directions of $SO(4,2)\sim SU(2,2)$. The four
directions which are orthogonal to the gauge group are non-compact
directions of $SU(2,1)\subset SU(2,2)$. The coset representative is
given by
\begin{eqnarray}
L&=&e^{a_1(c_{51}+c_{78})}e^{a_2(c_{36}+\tilde{c}_{53})}e^{a_3(c_{77}-c_{52})}e^{b_1(Y_1-Y_{23})}\nonumber
\\
&
&e^{-a_3(c_{77}-c_{52})}e^{-a_2(c_{36}+\tilde{c}_{53})}e^{-a_1(c_{51}+c_{78})}\,
.
\end{eqnarray}
We find the potential
\begin{equation}
V=-2g^2(5+3\cosh (2b_1))
\end{equation}
which, again, does not admit any non trivial critical points.
\\ \indent The next sector we will study is $SU(2)_{\textrm{diag}}$.
This symmetry is a diagonal subgroup of $SU(2)\times SU(2)$ in which
the first $SU(2)$ is a subgroup of $SU(4)\subset SU(4,2)$, and the
second $SU(2)$ is the $SU(2)$ factor in the gauge group. There are
four scalars in this sector. These scalars are non-compact
directions of $SU(2,1)$, and $L$ can be parametrized by
\begin{equation}
L=e^{a_1c_{10}}e^{a_2c_{14}}e^{a_3c_{15}}e^{b_1Y}e^{-a_3c_{15}}e^{-a_2c_{14}}e^{-a_1c_{10}}
\end{equation}
where
\begin{equation}
Y=Y_7-Y_6-Y_{12}-Y_{16}+Y_{17}+Y_{18}+Y_{30}+Y_{31}\, .
\end{equation}
The corresponding potential is
\begin{equation}
V=\frac{g^2}{8}[-101-28\cosh(4\sqrt{2}b_1)+\cosh(8\sqrt{2}b_1)]\, .
\end{equation}
\indent We now discuss its trivial critical point at $L=\mathbf{I}$.
This point is characterized by
\begin{equation}
V_0=-16g^2,\qquad
A_1=\textrm{diag}\left(-2,-2,-2,-2,-2,-2,2,2,2,2\right).
\end{equation}
The critical point has (6,4) supersymmetry. The gauge group is
broken down to its maximal compact subgroup $SU(4)\times SU(2)\times
 U(1)\times SU(2)$. The left handed supercharges transform as
 $\mathbf{6}$ under $SU(4)\sim SO(6)$ while the right handed
 supercharges transform as $\mathbf{4}$ under $SU(2)\times SU(2)\sim
 SO(4)$. So, the background isometry is given by $Osp(6|2, \mathbb{R})\times Osp(4|2,
 \mathbb{R})$. \\ \indent The non trivial critical point with $SU(2)_{\textrm{diag}}\times SU(2)\times SU(2)\times
 U(1)$ symmetry
 is given by
\begin{equation}
b_1=\frac{1}{\sqrt{2}}\cosh^{-1}\sqrt{\frac{3}{2}},\qquad
V_0=-25g^2\, .
\end{equation}
The associated $A_1$ tensor is given in appendix \ref{A1tensor}
equation \eqref{A1SU42_SU2} which can be diagonalized to
\begin{equation}
A_1=\textrm{diag}\left(\frac{11}{2},\frac{11}{2},\frac{11}{2},\frac{11}{2},-\frac{7}{2},
-\frac{7}{2},-\frac{5}{2},-\frac{5}{2},-\frac{5}{2},-\frac{5}{2}\right).
\end{equation}
So, this is a (4,0) point with
\begin{equation}
\frac{c_{(0)}}{c_{(1)}}=\frac{5}{4}\, .
\end{equation}

\section{Conclusions}\label{conclusion}
In this paper, we have studied critical points of $N=10$ three
dimensional gauged supergravity with both compact and non-compact
gauge groups. Remarkably, all critical points found in this paper
are AdS critical points. This is in contrast to the results of
\cite{N16Vacua} in which some Minkowski and dS vacua have been
found. All critical points found in this paper are listed in Table
1.
\\ \indent
The gauge groups considered in this work are only maximal subgroups
of $SO(10)\times U(1)$ and $E_{6(-14)}$. It is interesting to study
gaugings with other gauge groups which are not maximal subgroups of
$SO(10)\times U(1)$ and $E_{6(-14)}$ along with their scalar
potentials and the corresponding critical points. In particular,
non-semisimple gaugings are very interesting in the sense that they
are related to semisimple Yang-Mills gaugings which arise from
dimensional reductions of higher dimensional theories \cite{csym}.
Furthermore, studies of RG flows between critical points identified
in this work are of particular interest in studying deformations of
the dual two dimensional CFT's. We hope to give further results on
these issues in future works.
\\
\begin{tabular}{|c|c|c|c|c|}
  \hline
  Critical  & Gauge group & $V_0$ & Unbroken  & Unbroken  \\
  point & & & SUSY & gauge
  symmetry
  \\ \hline
  1 & $SO(10)\times U(1)$ & $-64g^2$ & $(10,0)$ & $SO(10)\times U(1)$ \\
  2 & $SO(10)\times U(1)$ & $-100g^2$ & - & $SO(7)$ \\
  3 & $SO(9)\times U(1)$ & $-64g^2$ & $(9,1)$ & $SO(9)\times U(1)$ \\
  4 &$SO(9)\times U(1)$ & $-\frac{1024}{9}g^2$ & $(2,1)$ & $G_2$ \\
  5 &$SO(9)\times U(1)$ & $-100g^2$ & $(1,0)$ & $G_2$ \\
  6 &$SO(8)\times SO(2)$ & $-64g^2$ & $(8,2)$ & $SO(8)\times SO(2)$ \\
  &  $\times U(1)$  &      &     &   $\times U(1)$    \\
  7 &$SO(8)\times SO(2)$ & $-100g^2$ & - & $SO(7)$ \\
  &  $\times U(1)$  &      &     &       \\
  8 &$SO(8)\times SO(2)$ & $-\frac{1024}{9}g^2$ & $(1,1)$ & $G_2$ \\
  &  $\times U(1)$  &      &     &       \\
  9 &$SO(7)\times SO(3)$ & $-64g^2$ & $(7,3)$ & $SO(7)\times SO(3)$ \\
  &  $\times U(1)$  &      &     &   $\times U(1)$    \\
  10 &$SO(7)\times SO(3)$ & $-\frac{1024}{9}g^2$ & $(0,1)$ & $G_2$ \\
  &  $\times U(1)$  &      &     &       \\
  11 &$SO(6)\times SO(4)$ & $-64g^2$ & $(6,4)$ & $SO(6)\times SO(4)$ \\
  &  $\times U(1)$  &      &     &   $\times U(1)$    \\
  12 &$SO(6)\times SO(4)$ & $-144g^2$ & $(0,2)$ & $SU(3)$ \\
  &  $\times U(1)$  &      &     &       \\
  13 &$SO(5)\times SO(5)$ & $-64g^2$ & $(5,5)$ & $SO(5)\times SO(5)$ \\
  14 &$SO(5)\times SO(5)$ & $-256g^2$ & $(3,0)$ & $SO(3)_{\textrm{diag}}$ \\
  15 &$G_{2(-14)}\times SU(2,1)$ & $-\frac{64}{9}g^2$ & $(7,3)$ & $G_{2(-14)}\times SU(2)$ \\
  & & & & $\times U(1)$ \\
  16 &$G_{2(-14)}\times SU(2,1)$ & $-\frac{100}{9}g^2$ & $(0,1)$ & $SU(3)$ \\
  17 &$G_{2(-14)}\times SU(2,1)$ & $-16g^2$ & $(2,3)$ & $SU(2)_{\textrm{diag}}\times U(1)$ \\
  18 &$F_{4(-20)}$ & $-16g^2$ & $(9,1)$ & $SO(9)$ \\
  19 &$F_{4(-20)}$ & $-25g^2$ & $(1,0)$ & $SO(7)$ \\
  20 &$SU(4,2)\times SU(2)$ & $-16g^2$ & $(6,4)$ & $SU(4)\times SU(2)$ \\
  & & & & $\times SU(2)\times U(1)$ \\
  21 &$SU(4,2)\times SU(2)$ & $-25g^2$ & $(4,0)$ & $SU(2)_{\textrm{diag}}\times SU(2)$ \\
  & & & & $\times SU(2)\times U(1)$ \\
  \hline
\end{tabular}\vspace{0.5 cm}\\
Table 1: Some critical points of $N=10$ gauged supergravity in three
dimensions.
\vspace{0.5 cm}
\\ \textbf{Acknowledgement}
We gratefully thank the Extreme Condition Physics Research
Laboratory at Department of Physics, Faculty of Science,
Chulalongkorn University for computing facilities. We also thank
Ahpisit Ungkitchanukit for reading the manuscript.

\appendix
\section{Essential formulae}\label{detail}
In this appendix, we give all necessary formulae in order to obtain
the scalar potential. We use the 52 generators of the $F_4$ subgroup
of $E_6$ from \cite{F4}. The remaining 26 generators are given in
\cite{E6}. The generators are normalized by
\begin{equation}
\textrm{Tr}(c_ic_j)=-6\delta_{ij}.
\end{equation}
With this normalization, we find that
\begin{eqnarray}
\mathcal{V}^{\alpha IJ}&=&-\frac{1}{6}\textrm{Tr}(L^{-1}T_G^\alpha
LX^{IJ}) \\
\mathcal{V}^{\alpha A}&=&\frac{1}{6}\textrm{Tr}(L^{-1}T_G^\alpha
LY^{A})\\
\mathcal{V}^{IJ}_{U(1)}&=&-\frac{1}{6}\textrm{Tr}(L^{-1}X
LX^{IJ}) \\
\mathcal{V}^A_{U(1)}&=&\frac{1}{6}\textrm{Tr}(L^{-1}X LY^{A})
\label{Vcoset}
\end{eqnarray}
where we have introduced the symbol $T^\alpha_G$ for gauge group
generators as in \cite{AP}. $T^\alpha_G$ will be replaced by some
appropriate generators of the gauge group being considered in each
gauging.
\\ \indent The following mapping provides the relation between $c_i$ and
$X^{IJ}$, generators of $SO(10)$,
\begin{eqnarray}
X^{1 2}&=& c_1,\,\, \, X^{1 3}= -c_2,\,\, \, X^{2 3}= c_3,\,\, \,
X^{3 4}=
c_6,\,\, \, X^{1 4}= c_4,\,\, \, X^{2 4}= -c_5,\nonumber \\
X^{1 5}&=& c_7,\,\, \, X^{2 5}= -c_8,\,\,\, X^{3 5}= c_9,\,\, \,
X^{4 5}= -c_{10},\,\, \, X^{5 6}= -c_{15},\,\, \, X^{1 6}= c_{11},
\nonumber \\
 X^{2  6}&=&
-c_{12},\,\, \, X^{4 6}= -c_{14},\,\, \, X^{3 6}= c_{13},\,\,\, X^{1
7}= c_{16},\,\, \, X^{27}= -c_{17},\,\, \, X^{4 7}= -c_{19}, \nonumber \\
X^{3 7}&=& c_{18},\,\, \, X^{6 7}= -c_{21},\,\, \, X^{ 5 7}=
-c_{20},\,\, \, X^{7 8}= -c_{36},\,\,\, X^{1 8}= c_{30},\,\, \, X^{2
8}= -c_{31},
\nonumber \\
X^{4 8}&=& -c_{33},\,\, \, X^{3 8}= c_{32},\,\, \, X^{6 8}=
-c_{35},\,\, \, X^{5 8}= -c_{34},\,\, \, X^{2 9}= -c_{46},\,\,\,
X^{1 9}= c_{45},
\nonumber \\
X^{4 9}&=& -c_{48},\,\,\, X^{3 9}= c_{47},\,\,\, X^{6 9}=
-c_{50},\,\, \, X^{5 9}= -c_{49}, X^{8 9}= -c_{52},\,\, \, X^{7 9}=
-c_{51},\nonumber \\
X^{1,10}&=&-c_{71},\,\,\,X^{2,10}=c_{72},\,\,\,X^{3,10}=-c_{73},\,\,\,X^{4,10}=c_{74},\,\,\,
X^{5,10}=c_{75},\nonumber
\\
X^{6,10}&=&c_{76},\,\,\,X^{7,10}=c_{77},\,\,\,X^{8,10}=c_{78},\,\,\,X^{9,10}=\tilde{c}_{53}\,
.
\end{eqnarray}
The $\tilde{c}_{53}$ and $\tilde{c}_{70}$ are defined by \cite{E6}
\begin{equation}
\tilde{c}_{53}=\frac{1}{2}c_{53}+\frac{\sqrt{3}}{2}c_{70} \qquad
\textrm{and}\qquad
\tilde{c}_{70}=-\frac{\sqrt{3}}{2}c_{53}+\frac{1}{2}c_{70}\, .
\end{equation}
All the $f^{IJ}$'s components can be obtained from the structure
constants of the $[X^{IJ},Y^A]$ given in \cite{F4} and \cite{E6}.
\\ \indent Generators of the $SO(p)\times SO(10-p)$ compact gauge group are given by
\begin{eqnarray}
T_1^{IJ}&=&X^{IJ},\qquad I, J=1,\ldots p,\nonumber \\
T_2^{IJ}&=&X^{IJ},\qquad I, J=p+1,\ldots 10\, .
\end{eqnarray}
The $U(1)$ subgroup is generated by $X=2\tilde{c}_{70}$.
\\ \indent In the non-compact $G_{2(-14)}\times SU(2,1)$ gauging,
the generators of $G_{2(-14)}$ can be obtained from combinations of
$SO(7)$ generators \cite{G2inSO(7)}
\begin{eqnarray}
T_1&=&\frac{1}{\sqrt{2}}(X^{36}+X^{41}),\,\,\,T_2=\frac{1}{\sqrt{2}}(X^{31}-X^{46}),\nonumber
\\
T_3&=&\frac{1}{\sqrt{2}}(X^{43}-X^{16}),\,\,\,T_4=\frac{1}{\sqrt{2}}(X^{73}-X^{24}),\nonumber
\\
T_5&=&-\frac{1}{\sqrt{2}}(X^{23}+X^{47}),\,\,\,T_6=-\frac{1}{\sqrt{2}}(X^{26}+X^{71}),\nonumber
\\T_7&=&\frac{1}{\sqrt{2}}(X^{76}-X^{21}),\,\,\, T_8=\frac{1}{\sqrt{6}}(X^{16}+X^{43}-2X^{72}),\nonumber
\\
T_9&=&-\frac{1}{\sqrt{6}}(X^{41}-X^{36}+2X^{25}),\,\,\,
T_{10}=-\frac{1}{\sqrt{6}}(X^{31}+X^{46}-2X^{57}),\nonumber
\\T_{11}&=&\frac{1}{\sqrt{6}}(X^{73}+X^{24}+2X^{15}),\,\,\,
T_{12}=-\frac{1}{\sqrt{6}}(X^{74}-X^{23}+2X^{65}),\nonumber
\\T_{13}&=&\frac{1}{\sqrt{6}}(X^{26}-X^{71}+2X^{35}),\,\,\,
T_{14}=\frac{1}{\sqrt{6}}(X^{21}+X^{76}-2X^{45}).
\end{eqnarray}
These generators are essentially the same as those used in
\cite{AP}, but we repeat them here for conveniences. The $SU(2,1)$
generators are given by
\begin{eqnarray}
J_1&=&-c_{52},\qquad J_2=-\tilde{c}_{53},\qquad J_3=-c_{78},\qquad
J_4=\tilde{c}_{70}, \nonumber \\
J_5&=&\frac{1}{\sqrt{2}}(Y_1+Y_6),\qquad
J_6=\frac{1}{\sqrt{2}}(Y_9+Y_{14}),\nonumber
\\J_7&=&\frac{1}{\sqrt{2}}(Y_{21}+Y_{24}),\qquad
J_8=\frac{1}{\sqrt{2}}(Y_{25}+Y_{30})\, .
\end{eqnarray}
We have normalized these generators according to the embedding
tensor given in section \ref{N10}.
\\ \indent In $SU(4,2)\times SU(2)$ gauging, the relevant generators
are given by
\begin{itemize}
  \item $SU(4,2)$:
  \begin{eqnarray}
  Q_i&=&c_i, \qquad i=1,\ldots , 15,\nonumber \\
  Q_{16}&=&\frac{1}{\sqrt{2}}(c_{52}+c_{77}),\,\,\,Q_{17}=\frac{1}{\sqrt{2}}(c_{51}-c_{78}),\,\,\,
  Q_{18}=\frac{1}{\sqrt{2}}(\tilde{c}_{53}-c_{36}),\nonumber \\ Q_{19}&=&\tilde{c}_{70},
  \,\,\,Q_{20}=\frac{1}{\sqrt{2}}(Y_{1}+Y_{23}),\,\,\,Q_{21}=\frac{1}{\sqrt{2}}(Y_{2}-Y_{22}),\nonumber
  \\
  Q_{22}&=&\frac{1}{\sqrt{2}}(Y_{3}+Y_{24}),\,\,\,Q_{23}=\frac{1}{\sqrt{2}}(Y_{4}-Y_{21}),\,\,\,
  Q_{24}=\frac{1}{\sqrt{2}}(Y_{5}+Y_{20}),\nonumber
  \\ Q_{25}&=&\frac{1}{\sqrt{2}}(Y_{6}+Y_{18}),\,\,\,
  Q_{26}=\frac{1}{\sqrt{2}}(Y_{7}-Y_{17}),\,\,\,Q_{27}=\frac{1}{\sqrt{2}}(Y_{8}-Y_{19}),\nonumber
  \\
  Q_{28}&=&\frac{1}{\sqrt{2}}(Y_{9}+Y_{27}),\,\,\,Q_{29}=\frac{1}{\sqrt{2}}(Y_{10}-Y_{29}),\,\,\,
  Q_{30}=\frac{1}{\sqrt{2}}(Y_{11}-Y_{25}),\nonumber
  \\ Q_{31}&=&\frac{1}{\sqrt{2}}(Y_{12}+Y_{30}),\,\,\,
  Q_{32}=\frac{1}{\sqrt{2}}(Y_{13}+Y_{26}),\,\,\,Q_{33}=\frac{1}{\sqrt{2}}(Y_{14}-Y_{28}),\nonumber
  \\
  Q_{34}&=&\frac{1}{\sqrt{2}}(Y_{15}-Y_{32}),\,\,\,Q_{35}=\frac{1}{\sqrt{2}}(Y_{16}+Y_{31})\, .
  \end{eqnarray}
  \item $SU(2)$:
  \begin{equation}
  K_1=\frac{1}{2}(c_{51}+c_{78}),\,\,\,
  K_2=-\frac{1}{2}(c_{52}-c_{77}),\,\,\,
  K_3=\frac{1}{2}(c_{36}+\tilde{c}_{53}).
  \end{equation}
\end{itemize}
To find the above generators, we first look at the generators of the
compact subgroup $SU(4)\times SU(2)\times U(1)$ of the $SU(4,2)$.
Using the fact that $SU(4)\sim SO(6)$ and $SU(2)\times SU(2)\sim
SO(4)$, we can identify $SU(4)\times SU(2)\times SU(2)$ with
$SO(6)\times SO(4)\subset SO(10)$. The $U(1)$ generator is simply
$\tilde{c}_{70}$.
\\ \indent The final non-compact gauge group is $F_{4(-20)}$. Its
generators can be easily identified by $c_1,\ldots, c_{52}$ in the
construction of the $E_6$ given in \cite{E6}.
\\ \indent
We can now compute the T-tensors using
\begin{eqnarray}
T^{IJ,KL}&=&\mathcal{V}^{IJ,\alpha} \mathcal{V}^{KL,\beta}
\delta^{SO(p)}_{\alpha\beta} -\mathcal{V}^{IJ,\alpha}
\mathcal{V}^{KL,\beta}
\delta^{SO(10-p)}_{\alpha\beta}+\frac{1}{3}(5-p)\mathcal{V}^{IJ}_{U(1)} \mathcal{V}^{KL}_{U(1)}, \\
T^{IJ,A}&=&\mathcal{V}^{IJ,\alpha} \mathcal{V}^{A,\beta}
\delta^{SO(p)}_{\alpha\beta} -\mathcal{V}^{IJ,\alpha}
\mathcal{V}^{A,\beta}
\delta^{SO(10-p)}_{\alpha\beta}+\frac{1}{3}(5-p)\mathcal{V}^{IJ}_{U(1)}
\mathcal{V}^A_{U(1)}
\end{eqnarray}
for compact gaugings and
\begin{eqnarray}
T^{IJ,KL}&=&\mathcal{V}^{IJ,\alpha} \mathcal{V}^{KL,\beta}
\eta^{G_1}_{\alpha\beta}
-K\mathcal{V}^{IJ,\alpha} \mathcal{V}^{KL,\beta} \eta^{G_2}_{\alpha\beta}, \\
T^{IJ,A}&=&\mathcal{V}^{IJ,\alpha} \mathcal{V}^{A,\beta}
\eta^{G_1}_{\alpha\beta} -K\mathcal{V}^{IJ,\alpha}
\mathcal{V}^{A,\beta} \eta^{G_2}_{\alpha\beta}
\end{eqnarray}
for non-compact gaugings with $K$ being $\frac{2}{3}$ and $6$ for
$G_1\times G_2$ being $G_{2(-14)}\times SU(2,1)$ and $SU(4,2)\times
SU(2)$, respectively. As in \cite{AP}, we use summation convention
over gauge indices $\alpha$, $\beta$ with the notation
$\delta^{G_0}$ and $\eta^{G_0}$ meaning that the summation is
restricted to the $G_0$ generators. For $F_{4(-20)}$ gauging, we
have the simpler expressions for the T-tensors namely
\begin{eqnarray}
T^{IJ,KL}&=&\mathcal{V}^{IJ,\alpha} \mathcal{V}^{KL,\beta}
\eta^{F_{4(-20)}}_{\alpha\beta},\nonumber \\
T^{IJ,A}&=&\mathcal{V}^{IJ,\alpha} \mathcal{V}^{A,\beta}
\eta^{F_{4(-20)}}_{\alpha\beta}\, .
\end{eqnarray}
\section{Explicit forms of the $A_1$ tensors}\label{A1tensor}
In this section, we give the explicit forms of the $A_1$ tensors
mentioned in the main text. We collect them here due to their
lengthly and complicated forms.
\begin{itemize}
 \item $SO(7)\times SO(3)\times U(1)$ gauging\\
 $G_2$ sector:
 \begin{equation}
A_1=\left(
      \begin{array}{cc}
        -8 \mathbf{I}_7 & 0 \\
        0 & M^{(1)}_3 \\
      \end{array}
    \right)\qquad M^{(1)}_3\left(
                             \begin{array}{ccc}
                               m_1 & m_4 & m_5 \\
                               m_4 & m_2 & m_6 \\
                               m_5 & m_6 & m_3 \\
                             \end{array}
                           \right)
    .\label{A1SO7}
\end{equation}
The elements of the matrix $M^{(1)}$ are given by
\begin{eqnarray}
m_1&=&\frac{1}{3} [21-\cos(2 a_3)-\cos(2 a_1) (1+3 \cos(2 a_3))+4
\cos(2 a_2) \sin^2(a_1) \sin^2a_3\nonumber \\ & &+4 \sin(2 a_1) \sin
a_2 \sin(2 a_3)]\nonumber
\\
m_2&=&-\frac{2}{3} [-11+\cos(2 a_2)-2 \cos^2a_2 \cos(2 a_3)]
\nonumber
\\
m_3&=&\frac{1}{3} [21-\cos(2 a_3)+\cos(2 a_1) (1+3 \cos(2 a_3))+4
\cos^2a_1  \cos(2 a_2) \sin^2a_3 \nonumber \\ & &-4 \sin(2 a_1) \sin
a_2 \sin(2 a_3)] \nonumber
\\
m_4&=&\frac{8}{3} \cos a_2 \sin a_3 [\cos a_1 \cos a_3-\sin a_1 \sin
a_2 \sin a_3] \nonumber
\\
m_5&=&\frac{1}{3} [[-2 \cos^2a_2+(-3+\cos(2 a_2)) \cos(2 a_3)]
\sin(2 a_1)\nonumber \\ & &-4 \cos(2 a_1) \sin a_2  \sin(2 a_3)]
\nonumber
\\
m_6&=&\frac{8}{3} \cos a_2  \sin a_3  (\cos a_3  \sin a_1 +\cos a_1
\sin a_2  \sin a_3 ).
\end{eqnarray}
  \item $G_{2(-14)}\times SU(2,1)$ gauging\\
  $SU(2)_{\textrm{diag}}$ sector:
  \begin{equation}
A_1=\left(
\begin{array}{cccccccccc}
 -4 & 0 & 0 & 0 & 0 & 0 & 0 & 0 & 0 & 0 \\
 0 & m_{22} & 0 & 0 & m_{52} & 0 & m_{72} & 0 & 0 & 0 \\
 0 & 0 & -4 & 0 & 0 & 0 & 0 & 0 & 0 & 0 \\
 0 & 0 & 0 & -4 & 0 & 0 & 0 & 0 & 0 & 0 \\
 0 & m_{52} & 0 & 0 & m_{55} & 0 & m_{75} & 0 & 0 & 0 \\
 0 & 0 & 0 & 0 & 0 & -4 & 0 & 0 & 0 & 0 \\
 0 & m_{72} & 0 & 0 & m_{75} & 0 & m_{77} & 0 & 0 & 0 \\
 0 & 0 & 0 & 0 & 0 & 0 & 0 & 2 & 0 & 0 \\
 0 & 0 & 0 & 0 & 0 & 0 & 0 & 0 & 2 & 0 \\
 0 & 0 & 0 & 0 & 0 & 0 & 0 & 0 & 0 & 2
\end{array}
\right)\label{G2SU2d}
\end{equation}
where
\begin{eqnarray}
m_{22}&=&\frac{1}{12} [-30+\cos[2 (a_1-a_2)]+\cos[2 (a_1+a_2)]-2
\cos(2 a_3)\nonumber \\& &+\cos(2 a_1) (2+6 \cos(2 a_3))+\cos(2 a_2)
(2-4 \cos^2a_1 \cos(2 a_3))\nonumber \\& &+8 \sin(2 a_1) \sin a_2
\sin(2
a_3)]\nonumber \\
m_{52}&=&\frac{1}{6} [(-2 \cos^2a_2+(-3+ \cos(2 a_2)) \cos(2 a_3))
\sin(2 a_1)\nonumber \\ & &+4 \cos(2 a_1) \sin a_2
\sin(2 a_3)]\nonumber \\
m_{72}&=&\frac{4}{3} \cos a_2  \sin a_3  (\cos a_3 \sin a_1 -\cos
a_1
 \sin a_2  \sin a_3 )\nonumber \\
m_{55}&=&\frac{1}{12} [-\cos[2 (a_1-a_2)] -\cos[2 (a_1+a_2)]-2
\cos(2 a_1)
 (1+3 \cos(2 a_3))\nonumber \\& &+\cos(2 a_2) (2-4
  \cos(2 a_3) \sin^2a_1 )-2 (15+\cos(2 a_3)\nonumber \\& &
  +4 \sin(2 a_1) \sin a_2  \sin(2 a_3))]\nonumber \\
m_{75}&=&\frac{4}{3} \cos a_2  \sin a_3  (\cos a_1 \cos a_3 +\sin
a_1
 \sin a_2  \sin a_3 )\nonumber \\
m_{77}&=&\frac{1}{3} [-7-\cos(2 a_2)+2 \cos^2a_2  \cos(2 a_3)].
\end{eqnarray}
\item $F_{4(-20)}$ gauging \\ $SU(3)$ sector:
\begin{equation}
A_1=\left(
\begin{array}{cccccccccc}
 -\frac{7}{2} & 0 & 0 & 0 & 0 & 0 & 0 & 0 & 0 & 0 \\
 0 & -\frac{7}{2} & 0 & 0 & 0 & 0 & 0 & 0 & 0 & 0 \\
 0 & 0 & -\frac{7}{2} & 0 & 0 & 0 & 0 & 0 & 0 & 0 \\
 0 & 0 & 0 & -\frac{7}{2} & 0 & 0 & 0 & 0 & 0 & 0 \\
 0 & 0 & 0 & 0 & a_{55} & 0 & 0 & a_{85} & a_{95} & 0 \\
 0 & 0 & 0 & 0 & 0 & -\frac{7}{2} & 0 & 0 & 0 & 0 \\
 0 & 0 & 0 & 0 & 0 & 0 & -\frac{7}{2} & 0 & 0 & 0 \\
 0 & 0 & 0 & 0 & a_{85} & 0 & 0 & a_{88} & a_{98} & 0 \\
 0 & 0 & 0 & 0 & a_{95} & 0 & 0 & a_{98} & a_{99} & 0 \\
 0 & 0 & 0 & 0 & 0 & 0 & 0 & 0 & 0 & \frac{11}{2}
\end{array}
\right)\label{A1F4SU3}
\end{equation}
where
\begin{eqnarray}
a_{55}&=&\frac{1}{16} [-50-2 \cos(2 a_1)+3 \cos[2 (a_1-a_3)]-8
\cos^2a_1  \cos(2 a_2) \cos^2a_3\nonumber \\ & &-2 \cos(2 a_3)+3
\cos[2 (a_1+a_3)]+8 \sin(2 a_1) \sin a_2  \sin(2
a_3)]\nonumber \\
a_{85}&=&\frac{1}{8} [[2 \cos^2a_2 +(-3+\cos(2 a_2)) \cos(2 a_3)]
\sin(2 a_1)\nonumber \\& &+4 \cos(2 a_1) \sin a_2  \sin(2
a_3)]\nonumber \\
a_{95}&=&\frac{1}{2} \cos a_2  [2 \cos a_1  \cos^2a_3 \sin a_2 +\sin
a_1  \sin(2 a_3
)]\nonumber \\
a_{88}&=&\frac{1}{16} [\cos[2 (a_1-a_2)]+\cos[2 (a_1+a_2)]+\cos(2
a_1) (2-6 \cos(2 a_3))\nonumber \\ & &-2 \cos(2 a_2) [1+2 \cos(2
a_3) \sin^2a_1]-2 (25+\cos(2 a_3)\nonumber \\& &+4 \sin(2 a_1) \sin
a_2 \sin(2 a_3))]\nonumber
\\
a_{99}&=&\frac{1}{4} [-13+\cos(2 a_2)+2 \cos^2a_2  \cos(2
a_3)]\nonumber \\
a_{98}&=&\frac{1}{2} \cos a_2  [-2 \cos^2a_3  \sin a_1  \sin a_2
+\cos a_1  \sin(2 a_3)].
\end{eqnarray}
\item $SU(4,2)\times SU(2)$ gauging \\ $SU(2)_{\textrm{diag}}$ sector:
\begin{equation}
A_1=\left(
\begin{array}{cccccccccc}
 -\frac{5}{2} & 0 & 0 & 0 & 0 & 0 & 0 & 0 & 0 & 0 \\
 0 & -\frac{5}{2} & 0 & 0 & 0 & 0 & 0 & 0 & 0 & 0 \\
 0 & 0 & -\frac{5}{2} & 0 & 0 & 0 & 0 & 0 & 0 & 0 \\
 0 & 0 & 0 & u_1 & u_4 & u_5 & 0 & 0 & 0 & 0 \\
 0 & 0 & 0 & u_4 & u_2 & u_6 & 0 & 0 & 0 & 0 \\
 0 & 0 & 0 & u_5 & u_6 & u_3 & 0 & 0 & 0 & 0 \\
 0 & 0 & 0 & 0 & 0 & 0 & \frac{11}{2} & 0 & 0 & 0 \\
 0 & 0 & 0 & 0 & 0 & 0 & 0 & \frac{11}{2} & 0 & 0 \\
 0 & 0 & 0 & 0 & 0 & 0 & 0 & 0 & \frac{11}{2} & 0 \\
 0 & 0 & 0 & 0 & 0 & 0 & 0 & 0 & 0 & \frac{11}{2}
\end{array}
\right)\label{A1SU42_SU2}
\end{equation}
where
\begin{eqnarray}
u_1&=&\frac{1}{16} [-50-\cos[2 (a_1-a_2)]-\cos[2 (a_1+a_2)]+2 \cos(2
a_3)\nonumber \\ & &-2 \cos(2 a_1) (1+3 \cos(2 a_3))+\cos(2 a_2)
(-2+4 \cos^2a_1 \cos(2 a_3))\nonumber \\ & &-8 \sin(2 a_1) \sin a_2
\sin(2
a_3)]\nonumber \\
u_2&=&\frac{1}{16} [-50+2 \cos(2 a_1)+\cos[2 (a_1-a_2)]-2 \cos(2
a_2)\nonumber \\ & &+\cos[2 (a_1+a_2)]+\cos(2 a_3) (2+6 \cos(2
a_1)+4 \cos(2 a_2) \sin^2a_1)\nonumber \\ & &+8 \sin(2 a_1) \sin a_2
\sin(2
a_3)]\nonumber \\
u_3&=&\frac{1}{4} [-13+\cos(2 a_2)-2 \cos^2a_2 \cos(2
a_3)]\nonumber \\
u_4&=&\frac{1}{8} [(2 \cos^2a_2-(-3+\cos(2 a_2)) \cos(2 a_3)) \sin(2
a_1)\nonumber \\ & &-4 \cos(2 a_1) \sin a_2  \sin(2
a_3)]\nonumber \\
u_5&=&\cos a_2  \sin a_3  (-\cos a_3  \sin a_1 +\cos a_1  \sin a_2
\sin a_3 )\nonumber
\\
u_6&=&-\cos a_2] \sin a_3  (\cos a_1  \cos a_3 +\sin a_1  \sin a_2
\sin a_3 )
\end{eqnarray}
\end{itemize}
\newpage
\section{Scalar potential for $SO(9)\times U(1)$ gauging in
$G_2$ sector}\label{SO9_potential}
\begin{eqnarray}
V&=&-\frac{1}{327680} g^2e^{-4 \sqrt{2} b_1} \bigg [-2 (4
(-1+e^{\sqrt{2} b_1})^3 (1+e^{\sqrt{2} b_1}) \cos[2 a_1] (1+3 \cos[2
a_3])\nonumber \\ & &+4 (-1+e^{2 \sqrt{2} b_1}) (29+6 e^{\sqrt{2}
b_1} +29 e^{2 \sqrt{2} b_1}-(-1+e^{\sqrt{2} b_1})^2 \cos[2
a_3]\nonumber \\ & &+4 (-1+e^{\sqrt{2} b_1})^2 (\cos[a_1]^2 \cos[2
a_2] \sin[a_3]^2-\sin[2 a_1] \sin[a_2] \sin[2 a_3])))^2\nonumber \\
& & +20 ((-1+e^{\sqrt{2} b_1})^4 (4 \cos[a_2]^2 \cos[2 a_3]+2 \cos[2
a_1] (-2 \cos[a_2]^2\nonumber \\ & & +(-3+\cos[2 a_2]) \cos[2
a_3])+8 \sin[2 a_1] \sin[a_2] \sin[2 a_3]))^2 \nonumber \\ &
&-2621440 e^{4 \sqrt{2} b_1} \cos[a_1]^2 \cos[a_2]^2 (\cos[a_3]
\sin[a_1]+\cos[a_1]
 \sin[a_2] \sin[a_3])^2 \times \nonumber \\ & & \sinh[\frac{b_1}{\sqrt{2}}]^6-384 e^{\sqrt{2} b_1}
 (-1+e^{\sqrt{2} b_1})^6 (4 \cos[2 a_3] \sin[2 a_1] \sin[a_2]\nonumber \\ & &+(3 \cos[2 a_1]-2
 \cos[a_1]^2 \cos[2 a_2]-1) \sin[2 a_3])^2\nonumber \\ & &-96 (-1+e^{2 \sqrt{2} b_1})^2
 (2 (4 (3+2 e^{\sqrt{2} b_1}+3 e^{2 \sqrt{2} b_1})+4 (-1+e^{\sqrt{2} b_1})^2 \cos[a_1]^2 \cos[a_3]^2
 \nonumber \\ & &+(-1+e^{\sqrt{2} b_1})^2 ((3+\cos[2 a_2]-2 \cos[2 a_1] \sin[a_2]^2) \sin[a_3]^2
 \nonumber \\ & &-2 \sin[2 a_1] \sin[a_2] \sin[2 a_3])))^2-4 (-1+e^{2 \sqrt{2} b_1})^2
 (2 (29-2 e^{\sqrt{2} b_1} (-3+\cos[2 a_2])\nonumber \\ & &+\cos[2 a_2]+e^{2 \sqrt{2} b_1}
 (29+\cos[2 a_2])+(e^{\sqrt{2} b_1}-1)^2 \cos[2 a_1] (2 \cos[a_2]^2\nonumber \\ & &-(\cos[2 a_2]-3)
  \cos[2 a_3])-2 (e^{\sqrt{2} b_1}-1)^2 (\cos[a_2]^2 \cos[2 a_3]\nonumber \\ & &+2 \sin[2 a_1] \sin[a_2]
  \sin[2 a_3])))^2-16 e^{\sqrt{2} b_1} (-1+e^{\sqrt{2} b_1})^6 (12 \cos[2 a_1] \sin[2 a_3]\nonumber \\ & &
  +16 \cos[2 a_3] \sin[2 a_1] \sin[a_2]-4 (1+2 \cos[a_1]^2 \cos[2 a_2]) \sin[2 a_3])^2\nonumber \\ & &
  -(-4 (-1+e^{\sqrt{2} b_1})^3 (1+e^{\sqrt{2} b_1}) \cos[2 a_1] (1+3 \cos[2 a_3])\nonumber \\ & &
  -4 (-1+e^{2 \sqrt{2} b_1}) (29+6 e^{\sqrt{2} b_1}+29 e^{2 \sqrt{2} b_1}-(-1+e^{\sqrt{2} b_1})^2
  \cos[2 a_3]\nonumber \\ & &+4 (-1+e^{\sqrt{2} b_1})^2 (\cos[a_1]^2 \cos[2 a_2] \sin[a_3]^2
  -\sin[2 a_1] \sin[a_2] \sin[2 a_3])))^2\bigg ]
\end{eqnarray}
\newpage
\section{Scalar potential for $SO(6)\times SO(4)\times U(1)$ gauging in
$SU(3)$ sector}\label{SO6potential}
\begin{displaymath}
\begin{array}{lcl}
V&=&-4 g^2 \bigg [\frac{1}{64}(-11+\cosh[2 b_1]-24 \cosh[b_1]
\cosh[b_2]+2 \cosh[b_1]^2 \cosh[2 b_2])^2 \\ & &-5 (\frac{1}{10}
(-3+\cosh[b_1] \cosh[b_2])^2(\sinh[b_1]^2+\cosh[b_1]^2 \sinh[b_2]^2)
\\ & &+\frac{1}{129600}(\frac{1}{2} (\frac{1}{48} (-48 \csc[a_4]
\sin[2a_4] (-\cosh[\frac{b_2}{2}] (\sin[a_1] \sin[a_3]\sin[a_5] \\ &
&+\cos[a_1] (\cos[a_2] \cos[a_5]\sin[a_4]+\cos[a_3]\sin[a_2]
\sin[a_5])) \sinh[\frac{b_1}{2}] \\ &
&-\cosh[\frac{b_1}{2}](\cos[a_6] (\cos[a_3]
\sin[a_1]-\cos[a_1]\sin[a_2] \sin[a_3])\sin[a_4]
\\ & &+(\cos[a_5] \sin[a_1]\sin[a_3]+\cos[a_1](\cos[a_3]
\cos[a_5]\sin[a_2] \\ & &-\cos[a_2] \sin[a_4]
\sin[a_5]))\sin[a_6])\sinh[\frac{b_2}{2}])(\cosh[\frac{b_1}{2}]
\cosh[\frac{b_2}{2}] (\cos[a_1]\times  \\ & &
\cos[a_2]\cos[a_6]+(\cos[a_3]\sin[a_1]-\cos[a_1] \sin[a_2]
\sin[a_3])\sin[a_5]\sin[a_6]) \\ & &+\cos[a_5](-\cos[a_3]
\sin[a_1]+\cos[a_1]\sin[a_2] \sin[a_3])\sinh[\frac{b_1}{2}]
\sinh[\frac{b_2}{2}]) \\ & &+48\csc[a_4] \sin[2 a_4]
(\cosh[\frac{b_2}{2}](\sin[a_1] \sin[a_3]\sin[a_5] \\ & &+\cos[a_1]
(\cos[a_2] \cos[a_5]\sin[a_4]+\cos[a_3] \sin[a_2]
\sin[a_5]))\sinh[\frac{b_1}{2}] \\ & &+\cosh[\frac{b_1}{2}]
(\cos[a_6] (\cos[a_3]\sin[a_1]-\cos[a_1] \sin[a_2]
\sin[a_3])\sin[a_4] \\ & &+(\cos[a_5]\sin[a_1]\sin[a_3]+\cos[a_1]
(\cos[a_3] \cos[a_5]\sin[a_2] \\ & &-\cos[a_2]\sin[a_4] \sin[a_5]))
\sin[a_6])\sinh[\frac{b_2}{2}])(\cosh[\frac{b_1}{2}]\cosh[\frac{b_2}{2}]
(\cos[a_1] \times  \\ & &
\cos[a_2]\cos[a_6]+(\cos[a_3]\sin[a_1]-\cos[a_1]\sin[a_2] \sin[a_3])
\sin[a_5]\sin[a_6]) \\ & &+\cos[a_5](-\cos[a_3]\sin[a_1]+\cos[a_1]
\sin[a_2] \sin[a_3])\sinh[\frac{b_1}{2}]\sinh[\frac{b_2}{2}]) \\ &
&-96 \cosh[\frac{b_2}{2}] \sinh[b_1]\cos[a_4]
\cos[a_5]\cosh[\frac{b_2}{2}] (\cos[a_3] \sin[a_1]
\\ & &
-\cos[a_1]\sin[a_2] \sin[a_3])(\cos[a_6]
(\cos[a_3]\sin[a_1]-\cos[a_1] \sin[a_2] \sin[a_3])\times \\ & &
\sin[a_4]
 +(\cos[a_5]\sin[a_1]\sin[a_3]+\cos[a_1](\cos[a_3]
\cos[a_5]\sin[a_2]-\cos[a_2]\times  \\ & & \sin[a_4]
\sin[a_5]))\sin[a_6])-192\cosh[\frac{b_1}{2}]^2\cosh[\frac{b_2}{2}]
\cos[a_4] (\cos[a_1] \cos[a_2]\times \\ & &\cos[a_6]
+(\cos[a_3]\sin[a_1]-\cos[a_1]\sin[a_2] \sin[a_3])\sin[a_5]
\sin[a_6]) \times  \\ & & (\cos[a_6]
(-\cos[a_3]\sin[a_1]+\cos[a_1]\sin[a_2] \sin[a_3])\sin[a_4]
\\ & &-(\cos[a_5] \sin[a_1]\sin[a_3]+\cos[a_1](\cos[a_3]
\cos[a_5]\sin[a_2]
\\ & &
-192\sinh[\frac{b_1}{2}]^2\sinh[\frac{b_2}{2}]\cos[a_4] \cos[a_5]
\cosh[\frac{b_2}{2}] (\cos[a_3]\sin[a_1] \\ & &-\cos[a_2]\sin[a_4]
\sin[a_5])) \sin[a_6])\sinh[\frac{b_2}{2}] -\cos[a_1]\sin[a_2]
\sin[a_3])(\sin[a_1]\times  \\ & & \sin[a_3]
\sin[a_5]+\cos[a_1](\cos[a_2] \cos[a_5]\sin[a_4]+\cos[a_3]\sin[a_2]
\sin[a_5]))  \\ & &+96\sinh[b_1]
\sinh[\frac{b_2}{2}]\cos[a_4](\sin[a_1]
\sin[a_3]\sin[a_5]+\cos[a_1]\times \\ & &(\cos[a_2] \cos[a_5]
\sin[a_4] +\cos[a_3]\sin[a_2] \sin[a_5]))(\cos[a_1]
\cos[a_2]\cos[a_6] \\ & &+(\cos[a_3] \sin[a_1]-\cos[a_1]\sin[a_2]
\sin[a_3])\sin[a_5] \sin[a_6])\sinh[\frac{b_2}{2}]
\\ & &+192
\cos[a_4](\cosh[\frac{b_2}{2}](\cos[a_2]
\cos[a_5]\sin[a_1]\sin[a_4]+(\cos[a_3] \sin[a_1]\sin[a_2]
\\ & &-\cos[a_1]\sin[a_3])
\sin[a_5])\sinh[\frac{b_1}{2}]+\cosh[\frac{b_1}{2}] (-\cos[a_6]
(\cos[a_1]\cos[a_3] \\ & &+\sin[a_1]\sin[a_2]
\sin[a_3])\sin[a_4]+(\cos[a_5](\cos[a_3] \sin[a_1] \sin[a_2]\\ & &
-\cos[a_1]\sin[a_3]) -\cos[a_2]\sin[a_1] \sin[a_4]\sin[a_5])
\sin[a_6])\sinh[\frac{b_2}{2}])\times  \\ & &
(\cosh[\frac{b_1}{2}]\cosh[\frac{b_2}{2}]
(\cos[a_2]\cos[a_6]\sin[a_1]-(\cos[a_1]\cos[a_3]\\
& &+\sin[a_1]\sin[a_2] \sin[a_3]) \sin[a_5]\sin[a_6])
+\cos[a_5](\cos[a_1]\cos[a_3]
\end{array}
\end{displaymath}
\begin{displaymath}
\begin{array}{lcl}
& &+\sin[a_1]\sin[a_2] \sin[a_3])\sinh[\frac{b_1}{2}]
\sinh[\frac{b_2}{2}])
+192\cos[a_4](\cos[a_5]\cosh[\frac{b_2}{2}]\times \\ & &
(\cos[a_1]\cos[a_3]+\sin[a_1] \sin[a_2]
\sin[a_3])\sinh[\frac{b_1}{2}] +\cosh[\frac{b_1}{2}]\times \\ &
&(\cos[a_2] \cos[a_6]
\sin[a_1]-(\cos[a_1]\cos[a_3]+\sin[a_1]\sin[a_2]\times
\\ & & \sin[a_3])\sin[a_5]
\sin[a_6])\sinh[\frac{b_2}{2}])
(-\cosh[\frac{b_1}{2}]\cosh[\frac{b_2}{2}]
(\cos[a_6](\cos[a_1]\cos[a_3]
\\ & &
+\sin[a_1]\sin[a_2] \sin[a_3]) \sin[a_4]+(-\cos[a_3]\cos[a_5]
\sin[a_1]\sin[a_2] \\ &
&+\cos[a_1]\cos[a_5]\sin[a_3]+\cos[a_2]\sin[a_1] \sin[a_4]
\sin[a_5])\sin[a_6]) \\ & &+(\cos[a_2]\cos[a_5]
\sin[a_1]\sin[a_4]+(\cos[a_3]\sin[a_1] \sin[a_2] \\ &
&-\cos[a_1]\sin[a_3])
\sin[a_5])\sinh[\frac{b_1}{2}]\sinh[\frac{b_2}{2}])
\\ & &
+96 (\sinh[\frac{b_1}{2}]\cos[a_2]
\cos[a_5]\cosh[\frac{b_2}{2}]\sin[a_3]-\cosh[\frac{b_1}{2}](\cos[a_6
\sin[a_2] \\ & &+\cos[a_2]\sin[a_3]
\sin[a_5]\sin[a_6])\sinh[\frac{b_2}{2}])(\cosh[\frac{b_1}{2}]\cosh[\frac{b_2}{2}]
(\sin[2 a_4] \sin[a_2]\times  \\ & & \sin[a_5]\sin[a_6]+\cos[a_2]
(-\sin[2a_4] \cos[a_6] \sin[a_3] \\ & &+2\cos[a_4]
\cos[a_3]\cos[a_5] \sin[a_6]))+(-\sin[2 a_4] \cos[a_5] \sin[a_2] \\
& &+2\cos[a_4] \cos[a_2]\cos[a_3]
\sin[a_5])\sinh[\frac{b_1}{2}]\sinh[\frac{b_2}{2}]))
\\
& & +\frac{1}{4} (4 \csc[a_4] \sin[2
a_4](\cosh[\frac{b_2}{2}](\cos[a_1] \cos[a_2]\cos[a_6] \\ &
&+(\cos[a_3]\sin[a_1]-\cos[a_1] \sin[a_2] \sin[a_3])\sin[a_5]
\sin[a_6])\sinh[\frac{b_1}{2}] \\ & &-\cos[a_1]\sin[a_2]
\sin[a_3])\sinh[\frac{b_2}{2}])(-\cosh[\frac{b_1}{2}]\cosh[\frac{b_2}{2}]
(\sin[a_1] \sin[a_3]\sin[a_5] \\ & &+\cos[a_1](\cos[a_2]
\cos[a_5]\sin[a_4]+\cos[a_3]\sin[a_2]\sin[a_5])) \\ &
&-(\cos[a_6](-\cos[a_3] \sin[a_1]+\cos[a_1]\sin[a_2]
\sin[a_3])\sin[a_4] \\ & &
-(\cos[a_5]\sin[a_1]\sin[a_3]+\cos[a_1](\cos[a_3] \cos[a_5]
\sin[a_2] \\ & &-\cos[a_2]\sin[a_4]
\sin[a_5]))\sin[a_6])\sinh[\frac{b_1}{2}]\sinh[\frac{b_2}{2}])
\\ & & +\cos[a_5]\cosh[\frac{b_1}{2}] (\cos[a_3] \sin[a_1] -4
\csc[a_4] \sin[2 a_4](\cosh[\frac{b_2}{2}](\cos[a_1] \cos[a_2]\times
 \\ & &\cos[a_6]+(\cos[a_3]\sin[a_1]-\cos[a_1]\sin[a_2]
\sin[a_3]) \sin[a_5] \sin[a_6])\sinh[\frac{b_1}{2}] \\ &
&+\cos[a_5]\cosh[\frac{b_1}{2}]
(\cos[a_3]\sin[a_1]-\cos[a_1]\sin[a_2] \sin[a_3])
\sinh[\frac{b_2}{2}])(\cosh[\frac{b_1}{2}]\times  \\ &
&\cosh[\frac{b_2}{2}] (\sin[a_1]\sin[a_3]\sin[a_5] +\cos[a_1]
(\cos[a_2] \cos[a_5]\sin[a_4] \\ &
&+\cos[a_3]\sin[a_2]\sin[a_5]))+(\cos[a_6](-\cos[a_3]\sin[a_1]+\cos[a_1]\sin[a_2]\times
 \\ & &\sin[a_3]) \sin[a_4]
-(\cos[a_5]\sin[a_1]\sin[a_3]+\cos[a_1](\cos[a_3] \cos[a_5]\sin[a_2]
\\ & &-\cos[a_2]\sin[a_4] \sin[a_5]))
\sin[a_6])\sinh[\frac{b_1}{2}]\sinh[\frac{b_2}{2}]) \\ &
&+(8\cosh[\frac{b_1}{2}]\cosh[\frac{b_2}{2}] \sin[2 a_4]\cos[a_2]
\cos[a_5] \sin[a_1] \\ & &+8 \sin[2
a_4]\sinh[\frac{b_1}{2}]\sinh[\frac{b_2}{2}]\cos[a_1]
\cos[a_3]\cos[a_6] \\ & &+8 \sin[2 a_4]
\sinh[\frac{b_1}{2}]\sinh[\frac{b_2}{2}]\cos[a_6] \sin[a_1]\sin[a_2]
\sin[a_3] \\ & &+16\cos[a_4] \cosh[\frac{b_1}{2}]
\cosh[\frac{b_2}{2}] \cos[a_3]\sin[a_1] \sin[a_2]\sin[a_5]
\\ & &-16 \cos[a_4]\sinh[\frac{b_1}{2}]\sinh[\frac{b_2}{2}]
\cos[a_3] \cos[a_5]\sin[a_1] \sin[a_2]\sin[a_6] \\ & &+8 \sin[2
a_4]\sinh[\frac{b_1}{2}]\sinh[\frac{b_2}{2}]\cos[a_2] \sin[a_1]
\sin[a_5]\sin[a_6] \\ & &-16
\cos[a_4]\cosh[\frac{b_1}{2}]\cosh[\frac{b_2}{2}] \cos[a_1]\sin[a_3]
\sin[a_5] \\ & &+16
\cos[a_4]\sinh[\frac{b_1}{2}]\sinh[\frac{b_2}{2}]\cos[a_1]
\cos[a_5]\sin[a_3]\sin[a_6])(\cosh[\frac{b_2}{2}]\times  \\
& & (-\cos[a_2]
\cos[a_6]\sin[a_1]+(\cos[a_1]\cos[a_3]+\sin[a_1]\sin[a_2]
\sin[a_3])\times  \\ & &\sin[a_5] \sin[a_6])
\sinh[\frac{b_1}{2}]+\cos[a_5]\cosh[\frac{b_1}{2}]
(\cos[a_1]\cos[a_3] +\sin[a_1]\times \\ & &\sin[a_2] \sin[a_3])
\sinh[\frac{b_2}{2}])+16\cos[a_4](\cosh[\frac{b_2}{2}](\cos[a_6]
(\cos[a_1]\cos[a_3] \\ & &+\sin[a_1] \sin[a_2]
\sin[a_3])\sin[a_4]+(-\cos[a_3]\cos[a_5] \sin[a_1]\sin[a_2]
\\ & &+\cos[a_1]\cos[a_5]\sin[a_3]+\cos[a_2] \sin[a_1]
\sin[a_4]\sin[a_5]) \sin[a_6])\sinh[\frac{b_1}{2}]
\end{array}
\end{displaymath}
\begin{equation}
\begin{array}{lcl}
& &+\cosh[\frac{b_1}{2}](\cos[a_2] \cos[a_5]
\sin[a_1]\sin[a_4]+(\cos[a_3]\sin[a_1]\sin[a_2] \nonumber \\ & &
-\cos[a_1]\sin[a_3]) \sin[a_5])\sinh[\frac{b_2}{2}])
(\cosh[\frac{b_1}{2}]\cos[a_5] \cosh[\frac{b_2}{2}]\times
\\ &
&(\cos[a_1]\cos[a_3]+\sin[a_1]\sin[a_2]\sin[a_3])-\sinh[\frac{b_1}{2}](\cos[a_2]\cos[a_6]
\sin[a_1] \\ & &-(\cos[a_1]\cos[a_3]+\sin[a_1]\sin[a_2]
\sin[a_3])\sin[a_5] \sin[a_6])\sinh[\frac{b_2}{2}])  \\ & &-16
\cos[a_4](\cosh[\frac{b_2}{2}](\cos[a_6]
(-\cos[a_3]\sin[a_1]+\cos[a_1]\sin[a_2] \sin[a_3])\times \\ &
&\sin[a_4] -(\cos[a_5] \sin[a_1]\sin[a_3]+\cos[a_1](\cos[a_3]
\cos[a_5]\sin[a_2] \\ & &-\cos[a_2]\sin[a_4]
\sin[a_5]))\sin[a_6])\sinh[\frac{b_1}{2}]+\cosh[\frac{b_1}{2}](\sin[a_1]
\sin[a_3]\sin[a_5] \\ & &+\cos[a_1](\cos[a_2]
\cos[a_5]\sin[a_4]+\cos[a_3]\sin[a_2]
\sin[a_5]))\sinh[\frac{b_2}{2}]) \times  \\ &
&(\cos[a_5]\cosh[\frac{b_1}{2}]\cosh[\frac{b_2}{2}]
(\cos[a_3]\sin[a_1]-\cos[a_1]\sin[a_2]\sin[a_3]) \\ & &+(\cos[a_3]
\sin[a_1] \sin[a_5]\sin[a_6]+\cos[a_1](\cos[a_2]\cos[a_6]
\\ & &-\sin[a_2]\sin[a_3] \sin[a_5]\sin[a_6]))
\sinh[\frac{b_1}{2}]\sinh[\frac{b_2}{2}])
+4(\cosh[\frac{b_2}{2}](\cos[a_6]\sin[a_2]\\ & & +\cos[a_2]\sin[a_3]
\sin[a_5] \sin[a_6])\sinh[\frac{b_1}{2}] +\cos[a_2]\cos[a_5]
\cosh[\frac{b_1}{2}]\sin[a_3]\times \\ & &\sinh[\frac{b_2}{2}])
(\cosh[\frac{b_1}{2}]\cosh[\frac{b_2}{2}] (-2 \sin[2a_4]\cos[a_5]
\sin[a_2] \\ & &+4\cos[a_4]\cos[a_2]\cos[a_3]
\sin[a_5])-\sinh[\frac{b_1}{2}](2\sin[2 a_4]
\sin[a_2]\sin[a_5]\sin[a_6] \\ & &+\cos[a_2] (-2 \sin[2a_4]\cos[a_6]
\sin[a_3]+4 \cos[a_4]\cos[a_3]\cos[a_5] \sin[a_6]))\times
 \\ &
&\sinh[\frac{b_2}{2}])
-8(\sinh[\frac{b_1}{2}]\cosh[\frac{b_2}{2}](\sin[2 a_4] \sin[a_2]
\sin[a_5]\sin[a_6]\\ & &+\cos[a_2] (-\sin[2a_4] \cos[a_6]\sin[a_3]
+2\cos[a_4]\cos[a_3] \cos[a_5]\sin[a_6]))\\ & &-
\cosh[\frac{b_1}{2}] (-\sin[2 a_4] \cos[a_5]\sin[a_2] +2
\cos[a_4]\cos[a_2] \cos[a_3]\sin[a_5])\times
\\ & &\sinh[\frac{b_2}{2}])
(\cos[a_2]
\cos[a_5]\cosh[\frac{b_1}{2}]\cosh[\frac{b_2}{2}]\sin[a_3]
+(\cos[a_6]\sin[a_2]\\ & &+\cos[a_2] \sin[a_3]
\sin[a_5]\sin[a_6])\sinh[\frac{b_1}{2}]\sinh[\frac{b_2}{2}]))))^2)\bigg
]
\end{array}
\end{equation}
\newpage
\section{Scalar potential for $SO(5)\times SO(5)$ gauging in $SO(3)_{\textrm{diag}}$
sector}\label{SO5potential}
\begin{displaymath}
\begin{array}{lcl}
V&=&-4g^2\bigg[4 (1+\cosh[2 b_1] \cosh[2 b_2])^2-\frac{1}{2}
(-1+\cosh[2 b_1] \cosh[2 b_2])^2 \times
\\ & &(1+\cosh[2 b_1] \cosh[2 b_2]) (\cos[a_2]^2 (\cos[2 a_3]+\cos[2 a_4]) \cos[a_5]^2
\\ & &-2 \sin[a_2]^2 \sin[a_5]^2+\sin[2 a_2] \sin[a_3] \sin[a_4] \sin[2 a_5]) \\ & &
+\frac{1}{64}(-1+\cosh[2 b_1] \cosh[2 b_2])^4(\cos[a_2]^2 (\cos[2
a_3] \\ & &+\cos[2 a_4]) \cos[a_5]^2-2 \sin[a_2]^2
\sin[a_5]^2+\sin[2 a_2] \sin[a_3]\times  \\ & & \sin[a_4] \sin[2
a_5])^2 -5 (\frac{33}{100} (\sinh[2 b_1]^2+\cosh[2 b_1]^2 \sinh[2
b_2]^2)
 \\ & &
-\frac{1}{6400}(-1+\cosh[2 b_1] \cosh[2 b_2]) (-41+23 \cosh[4 b_1]
\\ & &+2 \cosh[2 b_1]^2 (8 \cosh[4 b_2]+\cosh[8 b_2])) (\cos[2
(a_2-a_4)] \\ & &+\cos[2 (a_2+a_4)]+8 \cos[a_2]^2 \cos[2 a_3]
\cos[a_5]^2 \\ & &+\cos[2 a_4] (2+4 \cos[a_2]^2 \cos[2 a_5])-16
\sin[a_2]^2 \sin[a_5]^2 \\ & &+8 \sin[2 a_2]
 \sin[a_3] \sin[a_4] \sin[2 a_5])
 \\ & &
+\frac{1}{12800}(-1+\cosh[2 b_1] \cosh[2 b_2])^3 (1+\cosh[2 b_1]
\cosh[2 b_2]) \times  \\ & & (\cos[2 (a_2-a_4)]+\cos[2 (a_2+a_4)]+8
\cos[a_2]^2 \cos[2 a_3]\times  \\ & & \cos[a_5]^2+ \cos[2 a_4] (2+4
\cos[a_2]^2 \cos[2 a_5])-16 \sin[a_2]^2 \sin[a_5]^2 \\ & &+8 \sin[2
a_2] \sin[a_3]\times  \sin[a_4] \sin[2 a_5])^2 \\ & &
+\frac{1}{800}(\cos[a_2]^2 (5+\cos[2 a_3]+\cos[2 a_4]+(3+\cos[2 a_3]
\\ & &+\cos[2 a_4]) \cos[2 a_5])+8 \cos[2 a_5] \sin[a_2]^2-3
(-3+\cos[2 a_2])\times  \\ & & \sin[a_5]^2+4 \cosh[2 b_1]
 \cosh[2 b_2] (\cos[a_2]^2 (-\cos[a_4]^2 \\ & &+\cos[a_5]^2 \sin[a_3]^2)+(\cos[a_4]^2+\sin[a_2]^2
 \sin[a_4]^2) \sin[a_5]^2) \\ & &-2 (-1+\cosh[2 b_1] \cosh[2 b_2]) \sin[2 a_2] \sin[a_3] \sin[a_4]
 \times  \\ & & \sin[2 a_5]) \sinh[2 b_1]^2
+\frac{1}{800}\sinh[2 b_1]^2(\cos[a_1]^2 \cos[a_2]^2
\cos[a_4]^2\times  \\ & & (-1+\cosh[2 b_1] \cosh[2 b_2])+\cos[a_2]^2
\cos[a_4]^2 (-1+\cosh[2 b_1] \times
 \\ & &\cosh[2 b_2]) \sin[a_1]^2-\cos[a_4]^2 (3+\cosh[2 b_1] \cosh[2 b_2]) \sin[a_5]^2 \\ & &
 -(3+\cosh[2 b_1] \cosh[2 b_2]) (\cos[a_2] \cos[a_5] \sin[a_3] \\ & &-\sin[a_2] \sin[a_4]
 \sin[a_5])^2-2 (\cos[2 a_5] \sin[a_2]^2 \\ & &+\cos[a_2]^2 (\cos[2 a_3] \cos[a_5]^2-\cos[2 a_4]
 \sin[a_5]^2) \\ & &+\sin[2 a_2] \sin[a_3] \sin[a_4] \sin[2 a_5]))^2
\\ & &
+\frac{1}{3200}(-1+\cosh[2 b_1]^2 \cosh[4 b_2])(28+\cos[2 (a_2-a_3)]
\\ & &+\cos[2 (a_2+a_3)]+8 \cos[a_2]^2 \cos[2 a_4] \cos[a_5]^2+4
\cos[2 a_5] \\ & &+\cos[2 a_3] (2+4 \cos[a_2]^2 \cos[2 a_5])+8
\cos[2 a_2] \sin[a_5]^2
 \\ & &+16 \cosh[2 b_1] \cosh[2 b_2] (\cos[a_2]^2 (-\cos[a_4]^2+\cos[a_5]^2 \sin[a_3]^2) \\ & &+(\cos[a_4]^2
 +\sin[a_2]^2 \sin[a_4]^2) \sin[a_5]^2)-8 (-1+\cosh[2 b_1] \times  \\ & &\cosh[2 b_2]) \sin[2 a_2] \sin[a_3] \sin[a_4] \sin[2 a_5])
+\frac{1}{204800}(-1+\cosh[2 b_1]^2\times  \\ & & \cosh[4
b_2])(28+\cos[2 (a_2-a_3)]+\cos[2 (a_2+a_3)] +8 \cos[a_2]^2 \times
\\ & &\cos[2 a_4] \cos[a_5]^2+4 \cos[2 a_5]+\cos[2 a_3] (2+4
\cos[a_2]^2 \cos[2 a_5])
 \\ & &+8 \cos[2 a_2] \sin[a_5]^2+16 \cosh[2 b_1] \cosh[2 b_2] (\cos[a_2]^2 (-\cos[a_4]^2 \\ & &+\cos[a_5]^2
 \sin[a_3]^2)+(\cos[a_4]^2+\sin[a_2]^2 \sin[a_4]^2) \sin[a_5]^2)
\end{array}
\end{displaymath}
\begin{displaymath}
\begin{array}{lcl}
& & -8 (-1+\cosh[2 b_1] \cosh[2 b_2]) \sin[2 a_2]
 \sin[a_3] \sin[a_4] \sin[2 a_5])^2
\\ & &
+\frac{1}{400}(\sinh[2 b_1]^2+\cosh[2 b_1]^2 \sinh[2 b_2]^2)(4
\cos[a_2]^2 \cos[a_4]^2 \\ & &+ \cos[a_4]^2 (3+\cosh[2 b_1] \cosh[2
b_2]) \sin[a_2]^2+(3+\cosh[2 b_1]\times  \\ & & \cosh[2 b_2])
 \sin[a_4]^2-(-1+\cosh[2 b_1] \cosh[2 b_2]) (\cos[a_3]\times
\\ & &
\cos[a_5] \sin[a_1]-\cos[a_1] (\cos[a_5] \sin[a_2]
\sin[a_3]+\cos[a_2] \sin[a_4]\times  \\ & &
\sin[a_5]))^2-(-1+\cosh[2 b_1] \cosh[2 b_2]) (\cos[a_1] \cos[a_3]
\cos[a_5] \\ & &+\sin[a_1] (\cos[a_5] \sin[a_2] \sin[a_3]+\cos[a_2]
\sin[a_4] \sin[a_5]))^2)^2
 \\ & &
+\frac{1}{20} \cos[a_2]^2 \cos[a_4]^2 (-1+\cosh[2 b_1] \cosh[2
b_2])^3 (\cos[a_3]^2\times  \\ & &
 \sin[a_2]^2+(\cos[a_2] \cos[a_5] \sin[a_4]-\sin[a_2] \sin[a_3] \sin[a_5])^2)
 \\ & &
+\frac{3}{400}(-1+\cosh[2 b_1] \cosh[2 b_2])^2(- \cos[a_5]^2 \sin[2
a_2] \sin[a_3]\times  \\ & & \sin[a_4] ( \cosh[b_2] \sinh[b_1]
\cos[a_6]-\cosh[b_1] \sinh[b_2]\sin[a_6]) \\ & &+(\sin[2 a_2]
\sin[a_3] \sin[a_4] \sin[a_5]^2+\sin[a_2]^2 \sin[2 a_5]) \times  \\
& &(\cosh[b_2] \sinh[b_1] \cos[a_6]- \cosh[b_1] \sinh[b_2]
\sin[a_6]) \\ & &- \cos[a_3] \sin[2 a_2] \sin[a_4] \sin[a_5] (
\cosh[b_1] \sinh[b_2] \cos[a_6] \\ & &+\cosh[b_2] \sinh[b_1]
\sin[a_6])+ \cos[a_2]^2 \cos[a_5]\times  \\ & &
 (\cos[a_6] ( \cosh[b_1] \sinh[b_2] \sin[2 a_3]+\cosh[b_2] \sinh[b_1] (\cos[2 a_3] \\ & &+\cos[2 a_4]) \sin[a_5])
 +( \cosh[b_2] \sinh[b_1] \sin[2 a_3] \\ & &- \cosh[b_1] \sinh[b_2] (\cos[2 a_3]+\cos[2 a_4]) \sin[a_5]) \sin[a_6]))^2
 \\ & &
+\frac{3\text{  }}{6400}(-1+\cosh[2 b_1] \cosh[2 b_2])^2(\cos[a_3]
\sin[2 a_2] \sin[a_4]\times  \\ & & \sin[a_5] (4 \cosh[b_2]
\sinh[b_1] \cos[a_6]-4 \cosh[b_1] \sinh[b_2] \sin[a_6]) \\ &
&-\cos[a_5]^2 \sin[2 a_2]
 \sin[a_3] \sin[a_4] (4 \cosh[b_1] \sinh[b_2]\cos[a_6] \\ & &+4 \cosh[b_2] \sinh[b_1] \sin[a_6])+(\sin[2 a_2] \sin[a_3] \sin[a_4] \sin[a_5]^2
 \\ & &
+\sin[a_2]^2 \sin[2 a_5]) (4 \cosh[b_1] \sinh[b_2] \cos[a_6]+4
\cosh[b_2]\times  \\ & & \sinh[b_1] \sin[a_6]) +\cos[a_2]^2
\cos[a_5] (\cos[a_6] (-4 \cosh[b_2]\times  \\ & & \sinh[b_1] \sin[2
a_3]+4 \cosh[b_1] \sinh[b_2] (\cos[2 a_3]+\cos[2 a_4])\times  \\ & &
\sin[a_5])+(4 \cosh[b_1] \sinh[b_2] \sin[2 a_3]+4 \cosh[b_2]
\sinh[b_1] \times  \\ & & (\cos[2 a_3] +\cos[2 a_4]) \sin[a_5])
\sin[a_6]))^2
 \\ & &
+\frac{3}{400} (-1+\cosh[2 b_1] \cosh[2 b_2])^2(\cosh[b_2] (2
\cos[a_2] \cos[a_3]\times  \\ & & \cos[a_6](\cos[a_2] \cos[a_5]
\sin[a_3]-\sin[a_2] \sin[a_4] \sin[a_5]) \\ & &+(\cos[2 a_5] \sin[2
a_2] \sin[a_3] \sin[a_4]-2 \cos[a_5] (\cos[a_3]^2 \\ & &+\sin[a_2]^2
\sin[a_3]^2-\cos[a_2]^2 \sin[a_4]^2) \sin[a_5]) \sin[a_6])
\sinh[b_1] \\ & &+\cosh[b_1] (\cos[a_6] (\cos[2 a_5] \sin[2 a_2]
\sin[a_3] \sin[a_4] \\ & &-2 \cos[a_5] (\cos[a_3]^2+\sin[a_2]^2
\sin[a_3]^2 -\cos[a_2]^2 \sin[a_4]^2) \times  \\ & &\sin[a_5])+2
\cos[a_2] \cos[a_3] (-\cos[a_2] \cos[a_5] \sin[a_3] \\ & &
+\sin[a_2] \sin[a_4] \sin[a_5]) \sin[a_6]) \sinh[b_2])^2
\\ & &
+\frac{1}{50}(-1+\cosh[2 b_1] \cosh[2 b_2])^2( \cos[a_4]^2 \cos[a_5]
\sin[a_5]\times  \\ & & (-\cos[a_6] \cosh[b_2]
 \sinh[b_1]+\cosh[b_1] \sin[a_6] \sinh[b_2]) \\ & &-(\cos[a_2] \cos[a_5] \sin[a_3]-\sin[a_2] \sin[a_4] \sin[a_5]) (- \cosh[b_2]\times
\\ & &
(\cos[a_5] \cos[a_6]\sin[a_2] \sin[a_4]+\cos[a_2]\times  \\ & &
(\cos[a_6] \sin[a_3] \sin[a_5]-\cos[a_3] \sin[a_6]))
 \sinh[b_1]
\end{array}
\end{displaymath}
\begin{displaymath}
\begin{array}{lcl}
& &+ \cosh[b_1] (\cos[a_5] \sin[a_2] \sin[a_4]
\sin[a_6]+\cos[a_2]\times
\\ & &
(\cos[a_3] \cos[a_6]+\sin[a_3] \sin[a_5] \sin[a_6])) \sinh[b_2]))^2
 \\ & &
+\frac{1}{100} (-1+\cosh[2 b_1] \cosh[2 b_2])^2((\cos[a_2]^2
\cos[a_3] \cos[a_4]^2\times  \\ & & \cos[a_5] (\cosh[b_2] (\cos[a_3]
\cos[a_6] \sin[a_5]+\sin[a_3] \sin[a_6]) \times  \\ & &\sinh[b_1]
+\cosh[b_1] (\cos[a_6] \sin[a_3]-\cos[a_3] \sin[a_5]
\sin[a_6])\times  \\ & & \sinh[b_2])-(\cos[a_2] \cos[a_5] \sin[a_3]
\sin[a_4]-\sin[a_2] \sin[a_5])\times  \\ & & (\cosh[b_2] (\cos[a_5]
\cos[a_6] \sin[a_2]+\cos[a_2] \sin[a_4] \times  \\ & &(\cos[a_6]
\sin[a_3] \sin[a_5]-\cos[a_3] \sin[a_6])) \sinh[b_1] \\ &
&-\cosh[b_1] (\cos[a_5] \sin[a_2] \sin[a_6]+\cos[a_2] \sin[a_4]
\times  \\ & &(\cos[a_3] \cos[a_6]+\sin[a_3] \sin[a_5] \sin[a_6]))
\sinh[b_2])))^2
 \\ & &
+\frac{3}{400} (-1+\cosh[2 b_1] \cosh[2 b_2])^2(\cosh[b_2]
(\cos[a_6] \times   \\ & &(-\cos[2 a_5]
 \sin[2 a_2] \sin[a_3] \sin[a_4]+2 \cos[a_5] (\cos[a_3]^2
\\
& &+\sin[a_2]^2 \sin[a_3]^2-\cos[a_2]^2 \sin[a_4]^2) \sin[a_5])+2
\cos[a_2]\times  \\ & & \cos[a_3]
 (\cos[a_2] \cos[a_5] \sin[a_3]-\sin[a_2] \sin[a_4] \sin[a_5])\times  \\ & & \sin[a_6]) \sinh[b_1]
 +\cosh[b_1] (2 \cos[a_2] \cos[a_3] \cos[a_6]\times  \\ & & (\cos[a_2] \cos[a_5] \sin[a_3]-\sin[a_2]
  \sin[a_4] \sin[a_5]) \\ & &+(\cos[2 a_5] \sin[2 a_2] \sin[a_3] \sin[a_4]-2 \cos[a_5] (\cos[a_3]^2
  \\ & &+\sin[a_2]^2 \sin[a_3]^2-\cos[a_2]^2 \sin[a_4]^2) \sin[a_5]) \sin[a_6]) \sinh[b_2])^2
 \\ & &
+\frac{3}{400} (-1+\cosh[2 b_1] \cosh[2 b_2])^2(\cos[a_4]^2
\cos[a_5] \sin[a_5]\times  \\ & & ( \cosh[b_1] \sinh[b_2]
\cos[a_6]+\cosh[b_2] \sinh[b_1] \sin[a_6])
 \\ & &
-(\cos[a_2] \cos[a_5] \sin[a_3]-\sin[a_2] \sin[a_4] \sin[a_5])
(\cos[a_5]\times  \\ & & \sin[a_2] \sin[a_4] (\cosh[b_1]
\sinh[b_2]\cos[a_6]+\cosh[b_2]\times  \\ & &
\sinh[b_1]\sin[a_6])+\cos[a_2] (\cos[a_3](\cosh[b_2]
\sinh[b_1]\times  \\ & &  \cos[a_6]-\cosh[b_1]
\sinh[b_2]\sin[a_6])+\sin[a_3] \sin[a_5]\times  \\ & & (\cosh[b_1]
\sinh[b_2] \cos[a_6]+\cosh[b_2] \sinh[b_1] \sin[a_6]))))^2
 \\ & &
+\frac{1}{80} (-1+\cosh[2 b_1] \cosh[2 b_2])^2(-\cos[a_4]^2
\cos[a_5] \sin[a_5] \times  \\ & & (\cosh[b_2] \sin[a_6]
\sinh[b_1]+\cos[a_6] \cosh[b_1] \sinh[b_2]) \\ & &+(\cos[a_2]
\cos[a_5] \sin[a_3]-\sin[a_2] \sin[a_4] \sin[a_5]) ( \cosh[b_2]
\times  \\ & & (\cos[a_5] \sin[a_2] \sin[a_4]
 \sin[a_6]+\cos[a_2] (\cos[a_3] \cos[a_6] \\ & &+\sin[a_3] \sin[a_5] \sin[a_6])) \sinh[b_1]+ \cosh[b_1]
 (\cos[a_5] \cos[a_6] \sin[a_2] \sin[a_4] \\ & &+\cos[a_2] (\cos[a_6] \sin[a_3] \sin[a_5]-\cos[a_3] \sin[a_6])) \sinh[b_2]))^2
 \\ & &
+\frac{1}{100} (-1+\cosh[2 b_1] \cosh[2 b_2])^2 ((\cos[a_2]^2
\cos[a_3] \cos[a_4]^2 \cos[a_5]\times  \\ & & (\cosh[b_2]
(-\cos[a_6] \sin[a_3]+\cos[a_3] \sin[a_5] \sin[a_6]) \sinh[b_1] \\ &
&+\cosh[b_1] (\cos[a_3] \cos[a_6] \sin[a_5]+\sin[a_3] \sin[a_6])
\sinh[b_2]) \\ & &-(\cos[a_2] \cos[a_5] \sin[a_3]
\sin[a_4]-\sin[a_2]
 \sin[a_5]) (\cosh[b_2]\times  \\ & & (\cos[a_5] \sin[a_2] \sin[a_6]+\cos[a_2] \sin[a_4] (\cos[a_3] \cos[a_6]
 \\ & &+\sin[a_3] \sin[a_5] \sin[a_6])) \sinh[b_1]+\cosh[b_1] (\cos[a_5] \cos[a_6] \sin[a_2] \\ & &+\cos[a_2]
 \sin[a_4] (\cos[a_6] \sin[a_3] \sin[a_5]-\cos[a_3] \sin[a_6])) \sinh[b_2])))^2
 \\ & &
+\frac{1}{1600}(-1+\cosh[2 b_1] \cosh[2 b_2])^2( (\cosh[b_2]
(\cos[a_6] (-2 \cos[2 a_5] \times  \\ & &\sin[2 a_2] \sin[a_3]
\sin[a_4]+(1-\cos[2 a_2]+\cos[a_2]^2 (\cos[2 a_3]
\end{array}
\end{displaymath}
\begin{eqnarray}
& & +\cos[2 a_4])) \sin[2 a_5])+2 (\cos[a_2]^2 \cos[a_5] \sin[2 a_3]
\nonumber \\ & & -\cos[a_3] \sin[2 a_2] \sin[a_4] \sin[a_5])
\sin[a_6]) \sinh[b_1]+\cosh[b_1] \times \nonumber \\ & &(2 \cos[a_6]
(\cos[a_2]^2 \cos[a_5] \sin[2 a_3]-\cos[a_3] \sin[2 a_2]
\sin[a_4]\times \nonumber \\ & & \sin[a_5])+(2 \cos[2 a_5] \sin[2
a_2] \sin[a_3] \sin[a_4] -(\cos[a_2]^2\times \nonumber \\ & &
(\cos[2 a_3]+\cos[2 a_4])+2 \sin[a_2]^2) \sin[2 a_5]) \sin[a_6])
\sinh[b_2]))^2 \nonumber \\ & & +\frac{1}{1600}(-1+\cosh[2 b_1]
\cosh[2 b_2])^2((\cosh[b_2] (2 \cos[a_6] \times \nonumber \\ & &
(-\cos[a_2]^2 \cos[a_5] \sin[2 a_3]+\cos[a_3] \sin[2 a_2] \sin[a_4]
\sin[a_5])\nonumber \\ & & +(-2 \cos[2 a_5] \sin[2 a_2] \sin[a_3]
\sin[a_4]+(\cos[a_2]^2 (\cos[2 a_3]\nonumber \\ & &+\cos[2 a_4]) +2
\sin[a_2]^2) \sin[2 a_5]) \sin[a_6]) \sinh[b_1]+\cosh[b_1] \times
\nonumber \\ & &(\cos[a_6] (-2 \cos[2 a_5]
 \sin[2 a_2] \sin[a_3] \sin[a_4]+(1-\cos[2 a_2]\nonumber \\ & &+\cos[a_2]^2 (\cos[2 a_3]+\cos[2 a_4])) \sin[2 a_5])
 +2 (\cos[a_2]^2 \times \nonumber \\ & &\cos[a_5] \sin[2 a_3]-\cos[a_3] \sin[2 a_2] \sin[a_4] \sin[a_5])\sin[a_6]) \sinh[b_2]))^2)\bigg ]
\end{eqnarray}

\end{document}